\documentclass[APA,STIX1COL,hidelinks]{WileyNJD-v2}
\usepackage{adjustbox,booktabs,multirow}
\usepackage{caption}
\usepackage{subcaption}
\usepackage{enumitem}
\usepackage{bm}
\usepackage{xcolor,colortbl} 
\usepackage{scalerel}
\usepackage{tikz}
\usetikzlibrary{svg.path}
\definecolor{orcidlogocol}{HTML}{A6CE39}
\tikzset{
	orcidlogo/.pic={
		\fill[orcidlogocol] svg{M256,128c0,70.7-57.3,128-128,128C57.3,256,0,198.7,0,128C0,57.3,57.3,0,128,0C198.7,0,256,57.3,256,128z};
		\fill[white] svg{M86.3,186.2H70.9V79.1h15.4v48.4V186.2z}
		svg{M108.9,79.1h41.6c39.6,0,57,28.3,57,53.6c0,27.5-21.5,53.6-56.8,53.6h-41.8V79.1z M124.3,172.4h24.5c34.9,0,42.9-26.5,42.9-39.7c0-21.5-13.7-39.7-43.7-39.7h-23.7V172.4z}
		svg{M88.7,56.8c0,5.5-4.5,10.1-10.1,10.1c-5.6,0-10.1-4.6-10.1-10.1c0-5.6,4.5-10.1,10.1-10.1C84.2,46.7,88.7,51.3,88.7,56.8z};
	}
}

\newcommand\orcidicon[1]{\href{https://orcid.org/#1}{\mbox{\scalerel*{
				\begin{tikzpicture}[yscale=-1,transform shape]
				\pic{orcidlogo};
				\end{tikzpicture}
			}{|}}}}
\newcommand{\var}{\mathrm{Var}}

\makeatletter% 
\articletype{Research article}%

\received{ }
\revised{}
\accepted{}

\begin{document}
	\title{A product-limit estimator of the conditional survival function  when cure status is partially known}
	
	\author[1]{Wende C. Safari*}
	
	\author[2]{Ignacio L\'opez-de-Ullibarri}
	
	\author[3]{M. Amalia Jácome}
	
	\authormark{SAFARI \textsc{et al}}

	\address[1]{\orgdiv{Department of Mathematics,  Faculty of Computer Science}, \orgname{University of A Coru\~na}, CITIC,\orgaddress{\state{ A Coru\~na}, \country{Spain}}}
	
	\address[2]{\orgdiv{Department of Mathematics, Escuela Universitaria Polit\'ecnica}, \orgname{University of A Coru\~na}, \orgaddress{\state{Ferrol}, \country{Spain}}}
	
	\address[3]{\orgdiv{Department of Mathematics, Faculty of Science}, \orgname{University of A Coru\~na}, CITIC, \orgaddress{\state{A Coru\~na}, \country{Spain}}}
	
	\corres{*Wende C. Safari, Faculty of Computer Science, University of A Coru\~na, Elviña, 15071 A Coru\~na, Spain. \email{wende.safari@udc.es}}
	
	%\presentaddress{This is sample for present address text this is sample for present address text}
	
	\abstract[Abstract]{We introduce a nonparametric estimator of the conditional survival function in the mixture cure model for right censored data when cure status is partially known. The estimator is developed for the setting of a single continuous covariate but it can be extended to multiple covariates. It extends the estimator of \cite{beran1981nonparametric}, which ignores cure status information.   We obtain an almost sure representation, from which the strong consistency and asymptotic normality of the estimator are derived. Asymptotic expressions of the bias and variance demonstrate a reduction in the variance with respect to Beran's estimator. A simulation study shows that, if the bandwidth parameter is suitably chosen, our estimator performs better than others for an ample range of covariate values. A bootstrap bandwidth selector is proposed. Finally,  the proposed estimator is applied to a real dataset studying
		survival of sarcoma patients.}
	
	
	\jnlcitation{\cname{%
			\author{W. Safari}, 
			\author{I. L\'opez-de-Ullibarri}, 
			\author{M.A. J\'acome}  
		} (\cyear{2020}), 
		\ctitle{A product-limit estimator of the conditional survival function  when cure status is partially known}, \cjournal{Biometrical Journal}.}
	
	\maketitle

%==============================
\section{Introduction}\label{sec1:intro}
%==============================

The standard survival model assumes that, if there is no censoring, at some point all individuals will experience the event of interest. However, cure models have been developed because there are many situations where this assumption is not appropriate. In clinical settings, for example, it is very unlikely to have any recurrence of some tumors later than a certain period after radiation treatment. Such examples can be found in many other disciplines: some people will never get married, one-child mothers will never have a second child, some workers will never get a career shift, etc. In most literature, subjects in which an event will never take place are referred to as cured subjects. 

 The mixture cure model, originally proposed by \cite{boag1949maximum},  has received much attention in recent years. It assumes that the population is a mixture of cured and susceptible individuals. Note that here a ``cured'' individual is defined as being free of experiencing the event of interest, not necessarily cured in medical terms. The goal is to model the probability of cure and the  survival function of the uncured subjects, also called latency.  There has been substantial work on the mixture cure model, mostly with a (semi)parametric approach  \citep[see ][and references therein]{MallerZhou96,patilea2017general,amico2018cure}. These models are constructed under different (semi)parametric frameworks for the proportion of long-term survivors and/or the latency. However, when the underlying functions cannot be well approximated by the assumed (semi)parametric structures, applying those models will lead to biased estimates. Therefore, it is important to have completely nonparametric methods to model survival data with a cure fraction. \cite{maller1992estimating} proposed a consistent nonparametric estimator of the cure rate but their method cannot handle covariates.   Based on the estimator of the conditional survival function in \cite{beran1981nonparametric}, 
\cite{xu2014nonparametric} and \citeauthor{lopez2017nonparametric}  \citeyearpar{lopez2017nonparametric,lopez2017bnonparametric, lopez2020nonparametric} developed nonparametric methods for  the  mixture cure model in the presence of covariates.

Absence of individual's cure status (i.e., cured, uncured) is an important challenge for cure models.
 A subject that experiences the event is known to be uncured. However, censoring prevents from observing whether a censored subject would experience the event or not eventually. This hinders the classification of the censored observations in cured or uncured. In this situation, it is customary to assume no additional information on the cure status of the censored individuals, thus, to model the cure status as a latent variable. Nonetheless, there are situations where some of the censored individuals can be identified to be immune to the event of interest, that is, to be cured. For example, based on the result of a diagnosis procedure, some patients could be assumed to be cured from a given disease. Also, for some types of cancer it is extremely unlikely to have any recurrence later than a  given fixed time after treatment, known as a cure threshold. Another example of situation with individuals known to be cured is the analysis of hospital bed and intensive care unit (ICU) occupancy. In this, it is important to estimate the distribution of  time a patient will be in the hospital ward or ICU. Specifically,  modeling the time a patient stays in the hospital ward until admitted to the ICU. In the language of cure models, all patients who have died or have been discharged from the hospital bed without entering the ICU are censored and are known to be cured from the  ICU admission. This is of great interest to hospital management, particularly  in outbreaks of epidemic diseases such as the novel coronavirus disease (COVID-19).

Few authors have explored cure models when the cure status  is known for some censored observations.  \cite{laska1992nonparametric} and \cite{betensky2001nonparametric} discussed nonparametric cure rate estimation  with  cure status available, but  neither of them considered the presence of covariates.  \cite{nieto2008bayesian} proposed a Bayesian semiparametric approach for estimating a survival function with a cure fraction in the presence of covariates. A semiparametric approach based on a Cox proportional hazards cure model when cure information is partially known was studied by \cite{wu2014extension}.  \cite{bernhardt2016flexible} proposed a flexible cure rate model with potentially known cure threshold and  showed that ignoring a known cure threshold  may lead to biased estimates. 
Recently,  \cite{chen2018promotion} developed a nonparametric approach to modeling the covariate effects under the framework of promotion time. They considered a fixed cure threshold, so that observations censored at times larger than it are assumed to correspond to cured subjects.  Contrary to the  methods  mentioned, in this paper we develop    a completely nonparametric mixture cure model with covariates that can be applied in general situations, where the identification of the cured individuals does not  depend on a fixed cured threshold.  Examples of situations where a fixed cure threshold cannot be assumed were mentioned above: a study where a diagnostic procedure is used to discriminate between cured and uncured subjects, or a study of time to ICU admission of hospital inpatients, where discharge or death can occur before ICU admission.  Therefore, we propose a generalized product-limit estimator of the survival function that extends Beran's estimator when  cure status information is available.  From the proposed survival function estimator, further methods for the estimation of the cure rate and latency functions can be derived, in the spirit of \cite{xu2014nonparametric} and \citeauthor{lopez2017nonparametric} \citeyearpar{lopez2017nonparametric,lopez2017bnonparametric, lopez2020nonparametric}.

This paper is organized as follows. In Section \ref{sec2:curemodel},  after specifying the model notations, new estimators of the conditional cumulative hazard and survival functions are proposed, and some asymptotic results for them are given. For the choice of the bandwidth we propose a bootstrap  procedure in Section \ref{sec3:bandwidth}. In Section \ref{sec4:simulation}, we study the efficiency of the estimator of the survival function with a simulation study where our estimator is compared to  Beran's estimator, which ignores the available cure status information, as well as to the semiparametric estimator proposed by \cite{bernhardt2016flexible}.  In Section \ref{sec5:analysis}, the estimator is applied to estimate the distribution of the time to death from sarcoma cancer of $233$ patients from the University Hospital of Santiago de Compostela, Spain.  Section \ref{sec6:discussion} contains a discussion and thoughts for future  work.

%==============================
\section{Mixture cure model when cure status is partially known}\label{sec2:curemodel}
%==============================
\subsection{Model notation}

Let $Y$ be the survival time, $C$  the random censoring time and $\mathbf{X}$  a vector of covariates.  Assume that the survival time $Y$ is subject to random right censoring, so that instead of observing $Y$, only  $T=\min \left( Y,C\right) $ and $\delta=\textbf{1}(Y \leq C) $ can be observed.  The random variables $Y$ and $C$ are assumed to be conditionally independent given $\textbf{X}=\mathbf{x}$. Let $F(t| \mathbf{x})=P(Y\le t |  \mathbf{X}=\mathbf{x})$ denote the conditional distribution function of $Y$ and $G(t| \mathbf{x})=P(C\le t|\mathbf{X}=\mathbf{x})$ denote the conditional distribution function of $C$. It is assumed  that $X$, $Y$ and $C$ are  absolutely  continuous. We set $Y=\infty$ if the subject is cured. Let $\nu=\textbf{1}(Y=\infty)$ be an indicator of being cured. Note that $\nu$ is partially observed because $\delta=1$ implies $\nu=0$. In addition, when the cure status is partially known, $\nu=1$ is also observed for some censored individuals.   Suppose that  $\xi$   indicates whether the cure status is known ($\xi=1$) or not ($\xi=0$). Hence, the observations  $\{(\mathbf{X}_i,T_i,\delta_i,\xi_i,\xi_i\nu_i):i=1,\ldots,n\}$   can be classified into three  groups: (a) the individual is observed to have experienced the event  and therefore known to be uncured $ \left(\mathbf{X}_i,T_i=Y_i,\delta_i=1,\xi_i=1,\xi_i\nu_i=0\right)$; (b)  the lifetime is censored and the cure status is unknown $\left( \mathbf{X}_i,T_i=C_i,\delta_i=0,\xi_i=0,\xi_i\nu_i=0\right)$; and (c) the lifetime is censored and the individual is known to be cured $\left( \mathbf{X}_i,T_i=C_i,\delta_i=0,\xi_i=1,\xi_i\nu_i=1\right)$.   The probability of cure   is $1-p(\mathbf{x})=P(Y=\infty| \mathbf{X}=\mathbf{x})$, and the conditional survival function of the uncured individuals, also known as latency, is $S_0(t| \mathbf{x})=P(Y>t\mid Y<\infty, \mathbf{X}=\mathbf{x})$.
The mixture cure model writes the survival function  $S\left(t| \mathbf{x}\right)=1-F(t| \mathbf{x})=P(Y> t |  \mathbf{X}=\mathbf{x})$ as

\begin{align}
S\left(t\mid \mathbf{x}\right)=1-p(\mathbf{x})+p(\mathbf{x})S_0\left(t\mid \mathbf{x}\right).\label{eqn:mcm}
\end{align}
\noindent Assuming  model (\ref{eqn:mcm}),  the cure rate and the latency  can be written in terms of the survival function $S(t| \mathbf{x})$ as follows:
\begin{align*}
1-p(\mathbf{x})&=\lim\limits_{t \rightarrow \infty}S(t\mid \mathbf{x})> 0, \text{\ \ } S_0(t\mid \mathbf{x})=\frac{S(t\mid \mathbf{x}) -\{1-p(\mathbf{x})\}}{p(\mathbf{x})}.
\end{align*}

Therefore, the availability of a suitable estimator of  $S(t| \mathbf{x})$ would yield appropriate  estimators of  the cure probability and the latency directly.

 One key issue in cure models is identifiability.  This arises because of the lack of  cure status information at the end of the follow-up period, hence resulting in difficulties in distinguishing models with high incidence of susceptibles and long tails of the latency distribution from low incidence of susceptibles and short tails of the latency distribution \citep{li2001identifiability}.  Following the argumentation of \cite{hanin2014identifiability}, who discussed in detail  the identifiability of the mixture cure model,  model  (\ref{eqn:mcm}) is identifiable if the  latency function is proper. Thus, we assume that $\lim_{t\rightarrow\infty} S_0(t|\mathbf{x})=0$ for all $\textbf{x}$. This condition  is similar to the zero-tail constraint in  \cite{taylor1995semi}, \cite{lopez2017nonparametric}  and other papers. 
 
%--------------------------------------------------
\subsection{Proposed estimators}
%--------------------------------------------------
 Without loss of generality, for simplicity we only consider a single continuous covariate $X$  with density function $m(x)$.
As shown in  the Appendix, an estimator of the conditional cumulative hazard function of $Y$, $\Lambda(t| x)$, when  the cure  status is partially known is
 	\begin{equation}
 	\widehat{\Lambda }_{h}^c\left( t\mid x\right) =\sum_{i=1}^{n}\frac{\delta _{\lbrack i]}B_{h%
 			\left[ i\right] }\left( x\right) \textbf{1}\left(
 		T_{\left( i\right) }\leq t\right) }{\sum_{j=i}^{n}B_{h\left[
 			j\right] }\left( x\right) 
 		+\sum_{j=1}^{i-1}B_{h[j]}\left( x\right)\textbf{1}\left( \xi _{[j]}\nu _{[j]}=1\right) },\label{prop:lambda}
 	\end{equation}
 	where $X_{\left[ i\right] }$, $\delta _{\left[ i%
 		\right] }$, $\xi _{\left[ i\right] }$ and $\nu _{\left[ i\right] }$ are the  concomitants of the ordered observed
 	times $T_{\left( 1\right) }
 	\leq \cdot \cdot \cdot \leq T_{\left( n\right)
 	}$,  $	B_{h[i]}\left( x\right)$ are the Nadaraya-Watson weights,
 	\begin{align*}
 	B_{h[i]}\left( x\right)=\frac{K_h\left(x-X_{[i]}\right)}{\sum_{j=1}^{n}K_h\left(x-X_j\right)},
 	\end{align*} 	
 	 and $K_h(\cdot)=K(\cdot/h)/h$ is a kernel function $K(\cdot)$ rescaled with bandwidth $h$.   	 
 	  We  work with  Nadaraya-Watson kernel estimates since it is the natural choice for random design regression.

 	The corresponding product-limit estimator of the conditional survival function $S\left( t| x\right)$ when the cure status is partially known,  is 
 		\begin{align}
 	\widehat{S}_{h}^c\left( t\mid x\right)=\prod_{i=1}^{n}\left\{ 1-\frac{\delta
 		_{\lbrack i]}B_{h\left[ i\right] }\left( x\right) \textbf{1}\left( T_{\left( i\right)
 		}\leq t\right) }{\sum_{j=i}^{n}B_{h\left[
 		j\right] }\left( x\right) 
 	+\sum_{j=1}^{i-1}B_{h[j]}\left( x\right)\textbf{1}\left( \xi _{[j]}\nu _{[j]}=1\right)}\right\}.  \label{est_F}
 	\end{align}
An important feature of these estimators is that subjects who are known to be cured below time $T_{(i)}$ remain in the  risk set, i.e., they are counted in the denominator. In the following, we also refer to this estimator as $1-\widehat{F}_h^c(t| x)$. A motivation for the estimators (\ref{prop:lambda}) and  (\ref{est_F})  is given in the Appendix. 
\begin{proposition}\label{prop:property} The proposed estimator  $\widehat{S}_{h}^c\left( t| x\right)$  has  the following general properties.
	
\begin{enumerate}
	\item 	 When there are no censored observations known to be cured, i.e., $\xi_i\nu_i=0$ for  $i=1,\ldots,n$,  $\widehat{S}_{h}^c\left( t| x\right)$  reduces to  Beran's estimator:
	\begin{align}
	\widehat{S}_{h}\left( t\mid x\right)=\prod_{i=1}^{n}\left\{ 1-\frac{\delta
		_{\lbrack i]}B_{h\left[ i\right] }\left( x\right) \textbf{1}\left( T_{\left( i\right)
		}\leq t\right) }{\sum_{j=i}^{n}B_{h\left[
			j\right] }\left( x\right)}\right\}.   \label{est_Beran}
	\end{align}
	
	\item In the specific case when some individuals are observed as cured when their survival time exceeds a known fixed cure threshold,   $\widehat{S}_{h}^c\left( t| x\right)$ also reduces to Beran's estimator in (\ref{est_Beran}).
	\item When there is no censoring,   $\widehat{S}_{h}^c\left( t| x\right)$ reduces to the  kernel type  estimator  of the conditional survival function  \citep{nadaraya1964some}:
	
	\begin{align*}
	\widetilde{S}_{h}\left( t\mid x\right)=\sum_{i=1}^{n}B_{h\left[i\right] }\left( x\right) \textbf{1} \left( T_{\left( i\right)
	}> t\right). 
	\end{align*}%
	\item 
	In an unconditional setting, the proposed estimator is
	\begin{align*}
	\widehat{S}_{n}^c\left( t\right)=\prod_{i=1}^{n}\left\{ 1-\frac{\delta
		_{\lbrack i]} \textbf{1}\left( T_{\left( i\right)
		}\leq t\right) }{n-i+1+  \sum_{j=1}^{i-1}\textbf{1}\left( \xi _{[j]}\nu _{[j]}=1\right)}\right\}. 
	\end{align*}
In the particular case where an individual
is known to be cured only if the observed time is greater than a known fixed time, say $d$,  $\widehat{S}_{n}^c\left( t\right)$ reduces to the generalized maximum likelihood estimator in \cite{laska1992nonparametric}.
\end{enumerate}
\end{proposition}
\noindent The proof of these properties is outlined in the Appendix.
 \begin{proposition}\label{prop:npmle}
 The   $1-\widehat{F}_{h}^c\left( t|x\right)$ estimator in (\ref{est_F}) is the  nonparametric local maximum likelihood estimator of  $1-F(t| x)$.
 \end{proposition}

\noindent The proof of Proposition \ref{prop:npmle} is given in the Appendix.

%==============================
\subsection{Asymptotic results} \label{subsec:asymp}
%==============================
In this section, we investigate the asymptotic properties of  $\widehat{\Lambda}_h^c(t|x)$ and $\widehat{S}_{h}^c\left( t| x\right)$. In order to prove our asymptotic results, we consider  the following (sub)distribution functions:

\begin{align*}
H\left(t\mid x\right)&=P\left(T\leq t\mid X=x\right),\notag\\
H^1(t\mid x)   &=P\left(T\le t,\delta=1\mid X=x\right), \\
H^{11}(t\mid x)&=P\left(T\leq t,\xi=1,\nu=1\mid X=x \right),\notag\\
J(t\mid x)	   &=1-H\left(t\mid x\right)+H^{11}\left(t\mid x\right),
\end{align*}
\noindent  and  Assumptions \ref{ass:A1}--\ref{ass:A10}  stated in the Appendix. Assumptions like these have been  commonly used in  literature; see, e.g.,
    \cite{iglesias1999strong}.

Theorems \ref{thm:lambda} and \ref{thm:FhatFweigtht} below give  the asymptotic representations of $	\widehat{\Lambda }_{h}^c\left( t| x\right)$ and $1-	\widehat{F }_{h}^c\left( t|  x\right)$,  respectively.  Based on these results, in Corollary \ref{coro:supFhat} we show that  $	\widehat{\Lambda }_{h}^c\left( t|  x\right)$ and $1-	\widehat{F }_{h}^c\left( t| x\right)$  are  strongly consistent estimators of $	\Lambda \left( t|  x\right)$ and $1-F\left( t | x\right)$, respectively. The  asymptotic normality of $1-	\widehat{F }_{h}^c\left( t|  x\right)$ is proved in Theorem \ref{thm:asymnormality}. 

%-----------------------------------------
% THEOREM 1
%-----------------------------------------	
\begin{theorem} 	\label{thm:lambda} 
 Suppose that Assumptions \ref{ass:A1}--\ref{ass:A10}    hold, and	 the bandwidth $h=(h_n)$ satisfies $h\rightarrow 0, \log  n/nh\rightarrow 0 $ and $ nh^5/\log  n=O(1)$ as $n\rightarrow \infty$. Then, for $x\in I,t\in \lbrack a,b]$ we have 	
 
	\begin{equation*}
	\widehat{\Lambda }_{h}^c\left( t\mid  x\right) -\Lambda \left( t\mid  x\right)=\sum_{i=1}^{n}\widetilde{B}_{hi}\left( x\right) \zeta \left( T_{i},\delta_{i},\xi _{i},\nu _{i},t,x\right) +R_{n1}\left( t,x\right),
	\end{equation*}
 with
	\begin{align}
	\zeta \left( T_{i},\delta _{i},\xi _{i},\nu _{i},t,x\right) & =\frac{\mathbf{1}\left( T_{i}\leq t,\delta _{i}=1\right) }{J(T_{i}^-\mid  x)}   -\int_{0}^{t}\left\{ \mathbf{1}\left(T_{i}\geq v\right) +\mathbf{1}\left( T_{i}<v,\xi _{i}\nu _{i}=1\right) \right\}\frac{dH^{1}\left( v\mid  x\right) }{J^2\left( v^-\mid  x\right) },  \label{eqn:zeta}\\
	\widetilde{B}_{hi}\left( x\right) &=\frac{1}{m\left( x\right) }\frac{1}{nh}K\left( \frac{x-X_{i}}{h}\right), \label{def:Btilde}
	\end{align}
 where   $R_{n1}\left( t,x\right)$ satisfies $$
	\sup_{a\leq t\leq b,x\in I}\mid R_{n1}\left( t,x\right) \mid=O\left\{ \left( nh\right)^{-3/4}\left(\log n\right)^{3/4}\right\} {\text{ \ almost surely.\ }}$$
\end{theorem}

%-----------------------------------------
% THEOREM 2
%-----------------------------------------
\begin{theorem}
	\label{thm:FhatFweigtht}  	 Suppose that  Assumptions \ref{ass:A1}--\ref{ass:A10}   hold, and the	  bandwidth $h=(h_n)$ satisfies $h\rightarrow 0, \log  n/nh\rightarrow 0 $ and $ nh^5/\log  n=O(1)$ as $n\rightarrow \infty$. Then, for $x\in I,t\in \lbrack a,b]$ we have
	
	\begin{equation*}
	\widehat{F}_{h}^c\left( t\mid  x\right) -F\left( t\mid  x\right)
	=\left\{1-F\left( t\mid  x\right)\right\}\sum\limits_{i=1}^{n}\widetilde{B}_{hi}\left( x\right) 
	\zeta \left( T_{i},\delta _{i},\xi _{i},\nu _{i},t,x\right)
	+R_{n2}\left( t,x\right)
	\end{equation*}
where $	\zeta \left( T_{i},\delta _{i},\xi _{i},\nu _{i},t,x\right)$ is defined in (\ref{eqn:zeta}), $\widetilde{B}_{hi}(x)$ in (\ref{def:Btilde}) and $R_{n2}\left( t,x\right)$ satisfies
	\begin{equation}
		\sup_{a\leq t\leq b,x\in I}\mid R_{n2}\left( t,x\right) \mid=O\left\{ \left( nh\right)^{-3/4}\left(\log n\right)^{3/4}\right\}{\text{ almost surely.}} \label{eqn:Rn3}
	\end{equation}
\end{theorem}
\noindent The sketch of the proofs of Theorems \ref{thm:lambda} and \ref{thm:FhatFweigtht} is outlined in the Supplementary Material. The detailed proofs follow that of Theorem 2 of \cite{iglesias1999strong} for Beran's estimator.   As an immediate consequence  of these theorems,  the following corollary on the  strong consistency of the estimators $\widehat{\Lambda}_h^c(t|  x)$ and $1-\widehat{F}_h^c(t| x)$ is obtained.

%-----------------------------------------
% Corollary 1
%-----------------------------------------
\begin{corollary}
	\label{coro:supFhat} Suppose that Assumptions~\ref{ass:A1}--\ref{ass:A10}  hold, and the 	  bandwidth  $h=(h_n)$ satisfies $h\rightarrow 0, \log  n/nh\rightarrow 0 $ and $ nh^5/\log  n=O(1)$ as $n\rightarrow \infty$. Then, for $x\in I,t\in \lbrack a,b]$, we have
	 
	\begin{equation*}
	\sup_{a\leq t\le b ,x\in I}\mid \widehat{\Lambda}_{h}^c\left( t\mid x\right)
	-\Lambda\left( t\mid  x\right) \mid =O\left\{(nh)^{-1/2} \left(\log n\right)
	^{1/2}\right\} \text{\ \ almost surely,} \notag % \label{eqn:supLambda}
	\end{equation*}
and
	\begin{equation*}
	\sup_{a\leq t\le b ,x\in I}\mid \widehat{F}_{h}^c\left( t\mid  x\right)
	-F\left( t\mid  x\right) \mid =O\left\{(nh)^{-1/2} \left(\log n\right)
	^{1/2}\right\} {\text{ \ \ almost surely.}} %\label{eqn:supF}
	\end{equation*}
\end{corollary}
\noindent The proof of Corollary \ref{coro:supFhat} is outlined in the Appendix.

%----------------------------------------------
% Proposition 1
%---------------------------------------------
\begin{proposition}\label{prop:Bias_and_variance}
Suppose that Assumptions  \ref{ass:A1}--\ref{ass:A10}  hold, and the 	  bandwidth  $h=(h_n)$ satisfies $h\rightarrow 0, \log  n/nh\rightarrow 0 $ and $ nh^5/\log  n=O(1)$ as $n\rightarrow \infty$. Then, the  bias and  variance of $1-\widehat{F}_{h}^c\left( t|  x\right)$ are, respectively

	\begin{equation}
	\mu _{h,c}(t,x) = h^2 B_c(t,x) + O\left(h^4\right),
	\text{\ \ }
	\sigma^2_{h,c}(t,x) = (nh)^{-1} s_c^2(t,x)  +O(n^{-1}h), \label{eqn:biasvar}
	\end{equation}
		with
	\begin{align}
	B_{c}(t,x) &= \frac{\{1-F\left( t\mid x\right)\}\{2\Phi_c ^{\prime }\left( x,t,x\right) m^{\prime }\left( x\right) +\Phi_c
		^{\prime \prime }\left( x,t,x\right) m\left( x\right) \}d_K}{2m\left( x\right) }, \label{def:Bc} \\
	s_c^2(t,x) &=\frac{\{1-F\left( t\mid x\right)\}^{2}\Phi _{1}^c\left( x,t,x\right)c_K }{m\left( x\right) },\label{def:sc_2}
	\end{align}
 where $d_K=\int v^2 K(v)dv, c_{K}=\int K^{2}(v)dv$,
	\begin{align*}
	\Phi_c \left( y,t,x\right) =\text{E}\left\{ \zeta \left( T,\delta ,\xi , \nu 	,t,x\right) \mid X=y\right\}, \text{\ \ }
	\Phi_1^c\left(y,t,x\right) =\text{E}\left\{\zeta^2\left(T,\delta,\xi,\nu,t,x \right)\mid X=y\right\},
	\end{align*}
	with $\zeta \left( T,\delta ,\xi , \nu 	,t,x\right)$ given in (\ref{eqn:zeta}). Besides, $\Phi_c^\prime\left( y,t,x\right)$ and  $\Phi_c^{\prime\prime} \left( y,t,x\right)$ are the first and second derivatives of $\Phi_c \left( y,t,x\right)$ with respect to $y$.	
\end{proposition}
\noindent	The proof of Proposition \ref{prop:Bias_and_variance} is outlined in the Appendix. The following theorem, whose proof is in the Appendix, establishes the asymptotic normality of  $1-\widehat{F}_{h}^c\left( t\mid  x\right)$.

%----------------------------------------------
% Theorem 3
%---------------------------------------------
	\begin{theorem}
	\label{thm:asymnormality}
	Suppose that Assumptions \ref{ass:A1}--\ref{ass:A10}  hold are satisfied, then, for $x\in I$
	and  $t\in \lbrack a,b\rbrack$ it follows that:
	
	\begin{itemize}
		\item[(i)]  If $nh^{5}\rightarrow 0$ and $(\log  n)^{3}/nh\rightarrow 0,$ then $$(nh)^{1/2}\left\{ \widehat{F}_h^c(t\mid x)-F(t\mid x)\right\} \rightarrow	N(0,s_{c}^2(t,x)) \text{\ \ in distribution.}$$ 

	\item[(ii)] If $nh^{5}\rightarrow C^5>0$,  then $$(nh)^{1/2}\left\{ \widehat{F}_h^c(t\mid x)-F(t\mid x)\right\} \rightarrow
	N(C^{5/2}B_c(t,x),s_{c}^2(t,x)) \text{\ \ in distribution,}$$ 
\end{itemize}
with  $B_c(t,x)$ given in (\ref{def:Bc}), $s_{c}^2(t,x)$ in (\ref{def:sc_2}) and $C$ is  constant.
\end{theorem}

%--------------------------------------------
%       EFFECT OF IGNORING THE KNOWN CURES
%--------------------------------------------
%===============================================
\subsection{Effect of ignoring the cure status}\label{sec:eff}
%===============================================
In this section we make a theoretical comparison between the proposed estimator  $1-	\widehat{F }_{h}^c\left( t|  x\right)$ and Beran's estimator.  The asymptotic properties of Beran's estimator were obtained by \cite{iglesias1999strong} and \cite{van1997estimation}, among others. More precisely, in order to understand the effect of ignoring the cure status,  the dominant terms of the  bias and variance of  Beran's estimator are compared with  those of the proposed estimator.
 The asymptotic bias and variance of  Beran's estimator are, respectively,   

	\begin{equation}
\mu _{h}(t,x) = h^2 B(t,x) + O\left(h^4\right)
\text{\ and \ }
\sigma_{h}^2(t,x) = (nh)^{-1} s^2(t,x)  +O(n^{-1}h), \label{eqn:biasvarBeran}
\end{equation}
with

\begin{align}
B(t,x) &= \frac{\left\{1-F\left( t\mid x\right)\right\}\{2\Phi ^{\prime }\left( x,t,x\right) m^{\prime }\left( x\right) +\Phi
	^{\prime \prime }\left( x,t,x\right) m\left( x\right) \}d_K}{2m\left( x\right) }, \label{def:B} 
\end{align}
and
\begin{align}
s^2(t,x) &=\frac{\left\{1-F\left( t\mid x\right)\right\}^{2}\Phi _{1}\left( x,t,x\right)c_K }{m\left( x\right) },\label{def:s_2}
\end{align}
where, see  Lemmas 4 and 5 in \cite{lopez2017bnonparametric},
\begin{align}
\Phi \left( y,t,x\right) &=\int_{0}^{t}\frac{dH^{1}\left( v\mid  y\right) }{%
	1-H\left( v^-\mid  x\right) }-\int_{0}^{t}\frac{1-H\left( v^-\mid  y\right) }{%
	\left\{1-H\left( v^-\mid  x\right)\right\}^{2} }dH^{1}\left( v\mid  x\right), \label{def:Phi_Beran} \\%
\Phi _{1}\left( x,t,x\right) &=\int_{0}^{t}\frac{dH^{1}\left( v\mid x\right) }{%
	\left\{1-H\left( v^-\mid x\right)\right\}^{2} },\notag
\end{align}
and $\Phi^\prime(y,t,x)$ and  $\Phi^{\prime\prime}(y,t,x) $ are the first and the second derivatives of $\Phi \left( y,t,x\right)$ with respect to $y$. 
The expressions (\ref{eqn:biasvarBeran}) -- (\ref{def:s_2}) for Beran's estimator are equivalent to the bias and variance terms (\ref{eqn:biasvar})--(\ref{def:sc_2}) for $\widehat{S}_h^c(t \mid x)$, replacing $\Phi_c(x,t,x)$ and $\Phi_1^c(x,t,x)$ with $\Phi(x,t,x)$ and $\Phi_1(x,t,x)$, respectively. From Lemmas 2 and 5 in the Supplementary Material, we have
\begin{align*}
\Phi_c(y,t,x) =&\int_{0}^{t}\frac{dH^{1}\left( v\mid y\right) }{%
	1-H\left(v^-\mid x\right)+H^{11}\left(v^-\mid x\right) }-\int_{0}^{t}\frac{1-H\left(v^-\mid y\right)+H^{11}\left(v^-\mid y\right) }{%
	\left\{1-H\left(v^-\mid x\right)+H^{11}\left(v^-\mid x\right)\right\}^2 }dH^{1}\left( v\mid x\right),   \\
\Phi _{1}^c\left( x,t,x\right) =&\int_{0}^{t}\frac{dH^{1}\left( v\mid x\right) }{%
\left\{	1-H\left( v^-\mid x\right)+H^{11}(v^-\mid x)\right\}^2 }. 
\end{align*}
 As for the variance, when the cure status information is ignored then $H^{11}(t| x)=0$ for all $t$ and $x$. Therefore,  $\Phi_1^c(x,t,x) \leq \Phi_1(x,t,x) $. Notice that when the same bandwidth is used for both estimators, ignoring the cure status increases  asymptotically the variance of the estimator.

Returning to  the bias, by applying Lemma 3 in the Supplementary Material, we have 

\begin{align*}
\Phi _{c}^{\prime }\left( x,t,x\right) =\Phi ^{\prime }\left( x,t,x\right) &=-%
\frac{S^{\prime }\left( t^{-}\mid x\right) }{S\left( t^-\mid x\right) },
\end{align*}%

\noindent  where $S^{\prime }\left( t| x\right)$ is the derivative of $S\left( t| x\right)$ with respect to $x$,
meaning that the effect   of knowing the cure status on the bias is given by $\Phi_c^{\prime\prime}(x,t,x)$.  From Lemma 4 in the Supplementary Material,
\begin{align}	
\Phi _{c}^{\prime \prime }\left( x,t,x\right)&=2\int_{0}^{t}\frac{%
	G_{c}^{\prime }\left( v^{-}\mid x\right) }{1-G_{c}\left( v^{-}\mid x\right) }\frac{d }{ds}\left\{
\frac{S^{\prime }\left( s\mid x\right) }{S\left( s\mid x\right) }\right\}\Biggr|_{s=v^-} dv-\frac{%
	S^{\prime \prime }\left( t^-\mid x\right) }{S\left( t^-\mid x\right) }, \label{eqn:phic}
\end{align}
with%
\begin{equation*}
1-G_{c}(t\mid x)=1-G(t\mid x)+\pi _{1}(t,x)\{1-p(x)\}G_{1}(t\mid x), %\label{def:Gcb}
\end{equation*}%
where
\begin{align}
\pi _{1}\left( t,x\right)  =P\left( \xi =1\mid \nu =1,C\leq t,X=x\right),  \text{\  \ }
G_{1}\left( t\mid x\right)  =P\left( C\leq t\mid \nu =1,X=x\right) \label{eqn:piG}
\end{align}%

\noindent and $S^{\prime}(t| x)$, $S^{\prime\prime}(t| x)$ and $G^{\prime}(t \mid x)$ refer to the derivatives with respect to $x$. 
If the cure status is ignored, i.e., $\pi_{1}(x,t)=0$ for all $t$ and $x$, then (\ref{eqn:phic}) reduces to 
\begin{equation*}
\Phi ^{\prime \prime }\left( x,t,x\right) =2\int_{0}^{t}\frac{G^{\prime
	}\left( v^{-}\mid x\right) }{1-G\left( v^{-}\mid x\right) }\frac{d}{d s}\left\{ \frac{S^{\prime
	}\left( s\mid x\right) }{S\left( s\mid x\right) }\right\}\bigg\vert_{s=v^-} dv-\frac{S^{\prime \prime
	}\left( t^-\mid x\right) }{S\left( t^-\mid x\right) }.
\end{equation*}
In terms of bias, the advantage of knowing the cure status  is not straightforward as it depends on the derivative with respect to $x$ of the cure probability $1-p(x)$ and the functions  $\pi_1(t,x)$ and $G_1(t,x)$ in (\ref{eqn:piG}). This implies that there is  no  guarantee that   there will  be  a gain in terms of bias for the proposed estimator with respect to Beran's estimator.

%===============================================
\section{Bandwidth selection}\label{sec3:bandwidth}
%===============================================
Bootstrap procedures have been successfully used to address the issue of bandwidth selection in the context of the mixture cure model (\citeauthor{lopez2017nonparametric}, \citeyear{lopez2017nonparametric,lopez2017bnonparametric}). Next,  we propose a bootstrap bandwidth selector to choose the smoothing parameter $h$ of the proposed estimator $\widehat{S}_h^c(t|x)$.    The bootstrap bandwidth, $h_x^*$, is the bandwidth minimizing the bootstrap version of the mean integrated squared error (MISE). This bootstrap MISE  can be approximated using Monte Carlo by:
\begin{equation}
\text{MISE}_{x}^{\ast}(h)\simeq\dfrac{1}{B}\sum_{b=1}^{B}\int \left\{\widehat{S}_{h}^{c,\ast b}(v\mid x)-\widehat{S}_{g_x}^c(v\mid x) \right\}^2\omega(v,x) dv,
\label{eq:MSEb}
\end{equation}	
where $\widehat{S}_{h}^{c,\ast b}( t|x)$ is the proposed estimator computed with the $b$th bootstrap resample and a bandwidth $h$, and $\widehat{S}_{g_x}^c( t| x)$ is the same estimator computed with  the original sample  and with a pilot bandwidth $g_x$. Note that $\omega(v,x)$ is a nonnegative weight function, intended to give lower weight in the right tail of the distribution. The algorithm to compute the bootstrap bandwidth  for a fixed covariate value $x$, is as follows:
\begin{step}
\normalfont	With the original sample and the pilot bandwidth $g_x$, compute $\widehat S_{g_x}^c(t| x)$.
\end{step}
\begin{step}
\normalfont	Choose a dense enough grid of $L$ bandwidths $\{h_1,\ldots,h_L\}$.
\end{step}
\begin{step} \label{step3}
\normalfont Generate $B$ bootstrap resamples  $ \{ (X_i^{(b)}, T_i^{*(b)}, \delta_i^{*(b)},  \xi_i^{*(b)},\xi_i^{*(b)}\nu_i^{*(b)} ): i=1,\ldots,n  \}$, for $b=1, \dots, B$.
\end{step}
\begin{step}
\normalfont For the $b$th bootstrap resample and the bandwidths $h_l$, for $l=1,\ldots,L$, compute  $\widehat{S}_{h_l}^{c,\ast b}(t| x)$.
\end{step}

\begin{step}
\normalfont	For $h_l, l=1,\dots,L$, compute the Monte Carlo approximation of MISE$_{x}^*(h_l)$ given by (\ref{eq:MSEb}).
\end{step}
\begin{step}
\normalfont	The bootstrap bandwidth, $h_x^*$, is the  bandwidth of the grid  $\{h_1,\ldots,h_L\}$ that minimizes the approximation of $\text{MISE}_{x}^{\ast}(h)$ in (\ref{eq:MSEb}).
\end{step}
The bootstrap resamples in step \ref{step3} are generated as follows: fix $x$, for  $i=1,\ldots,n,$ set $X^*_i=X_i$ and generate a 4-tuple $(T_i^*,\delta_i^*,\xi_{i}^*,\xi_{i}^*\nu_{i}^*)$ from the weighted empirical conditional distribution of  $\{(T_1,\delta_1,\xi_{1},\xi_1 \nu_1),\ldots,(T_n,\delta_n,\xi_{n},\xi_n \nu_n)\}$:

$$
\widehat{F}_{g_x}(t,d,w,z\mid x)=\sum_{i=1}^{n}B_{g_xi}(x)\bm{1}\left(T_i\le t,\delta_{i}\le d,\xi_i\le w, \xi_{i}\nu_{i}\le z\right)
$$
where $B_{g_xi}(x)$ are the  Nadaraya-Watson weights with   bandwidth $g_x$.

The pilot bandwidth $g_x$ should tend to $0$ at a slower rate than $h_x^*$. This oversmoothing pilot bandwidth
is required for the bootstrap integrated squared bias and  variance to be asymptotically
efficient estimators of the integrated squared bias and  variance terms. For practical applications we recommend to use $g_x= c_x n^{-1/9}$, as suggested by \cite{lidatta2001bootstrap}, which coincides with the optimal order obtained by \cite{CaoGM1993bootstrap} for the uncensored case. Simulation results in the Supplementary Material (see also \citeauthor{lopez2017nonparametric}, \citeyear{lopez2017nonparametric, lopez2017bnonparametric}) show that the choice of the pilot bandwidth has a small effect on the selected bootstrap bandwidth. We propose to use the same local pilot bandwidth as in \citeauthor{lopez2017nonparametric} \citeyearpar{lopez2017nonparametric, lopez2017bnonparametric}:

\begin{equation*}
g_x = \frac{d_{k}^+(x) + d_{k}^-(x)}{2}100^{1/9} n^{-1/9},
\end{equation*}
where $d_k^+(x)$ and $d_k^-(x)$ are the distances  from $x$ to the $k$th nearest neighbor on the right and left, and $k$ is a suitably chosen integer depending on the sample size. If there are not at least $k$ neighbors on the right (or left), we use $d_k^+(x)=d_k^-(x)$ (or $d_k^-(x)=d_k^+(x)$).  Following  \citeauthor{lopez2017nonparametric} \citeyearpar{lopez2017nonparametric, lopez2017bnonparametric}, we suggest setting $k = [n/4]$.

%=================================
\section{Simulation study}\label{sec4:simulation}
%=================================
We studied the practical performance of $\widehat{S}_h^c(t| x)$ through a simulation study.  We considered the conditional survival function $ S(t| x) = 1-p(x) + p(x) S_0(t| x)$ where 
\begin{equation*}
S_{0}\left( t\mid x\right) =\left\{
\begin{array}{ll}
\dfrac{\exp \left( -\alpha \left( x\right) t\right) -\exp \left( -\alpha
	\left( x\right) 4.605\right) }{1-\exp \left( -\alpha \left( x\right)
	4.605\right) } & 0\leq t\leq 4.605 \\
0 & t>4.605
\end{array}%
\right., \ \ \text{ \  \ }\alpha \left( x\right) =\exp \left( \frac{x+20}{40}\right).
\end{equation*}
 We simulated  two scenarios given by the cure rates:
 \begin{equation*}
 1-p_1(x)=1-\frac{\exp \left( 0.476+0.358x\right) }{1+\exp \left(
	0.476+0.358x\right) }, \ \ \text{ \  \ } 1-p_2(x)=0.5-\frac{1}{16000}x^{3}.
\end{equation*}

 The censoring variable $C$ was generated from an exponential distribution with mean  $10/3$.  The covariate $X$ was  uniformly distributed on the interval $[-20,20]$. The  percentage of censoring was $54\%$ and the average cure probability  $0.467$  in Scenario 1, whereas in Scenario 2,  $61\%$ of the observations were censored and the average  cure probability was $0.5$.  In both scenarios,  the proportion of the identified cured individuals  was $\pi=0.2, 0.8$  and $1$. Data were generated so that the censoring times $C$ and the lifetimes $Y$ were independent conditionally on $X$. We generated $1000$ datasets of sample sizes $n=50, 100$ and $200$. This section contains the results for $\pi = 0.8$ and $n = 100$; the rest of the results can  be found in the Supplementary Material.
 
 Our first goal was to evaluate the performance of  $\widehat S_h^c(t| x)$ in terms of the MISE. It was approximated over a grid of $100$ bandwidths equispaced in a logarithmic scale, from $h_1=3$ to $h_{100}=20$ in Scenario 1, and from $h_1=4$ to $h_{100}=100$ in Scenario 2. For the weight function we chose $\omega(t,x)=\bm{1}(a_x\le t \le b_x)$ where $a_x=0$ and  $b_x=\tau_{x}$, the $90$th percentile of  $S_0(t| x)$.   We compared $\widehat S_h^c(t| x)$ computed in a grid of bandwidths  with  Beran's estimator, $\widehat S_h(t| x)$,  computed with the optimal bandwidth. The semiparametric estimator by \cite{bernhardt2016flexible},  which fits a logistic regression for the cure probability and seminonparametric accelerated failure time model for the latency function, was also considered for comparison. The semiparametric estimator is expected to perform well in Scenario 1. We chose the Epanechnikov kernel to compute $\widehat S_h^c(t| x)$ and $\widehat S_h(t| x)$.

\begin{figure}[htbp]	
	\begin{subfigure}[b]{.5\textwidth}
	\includegraphics[width=\textwidth]{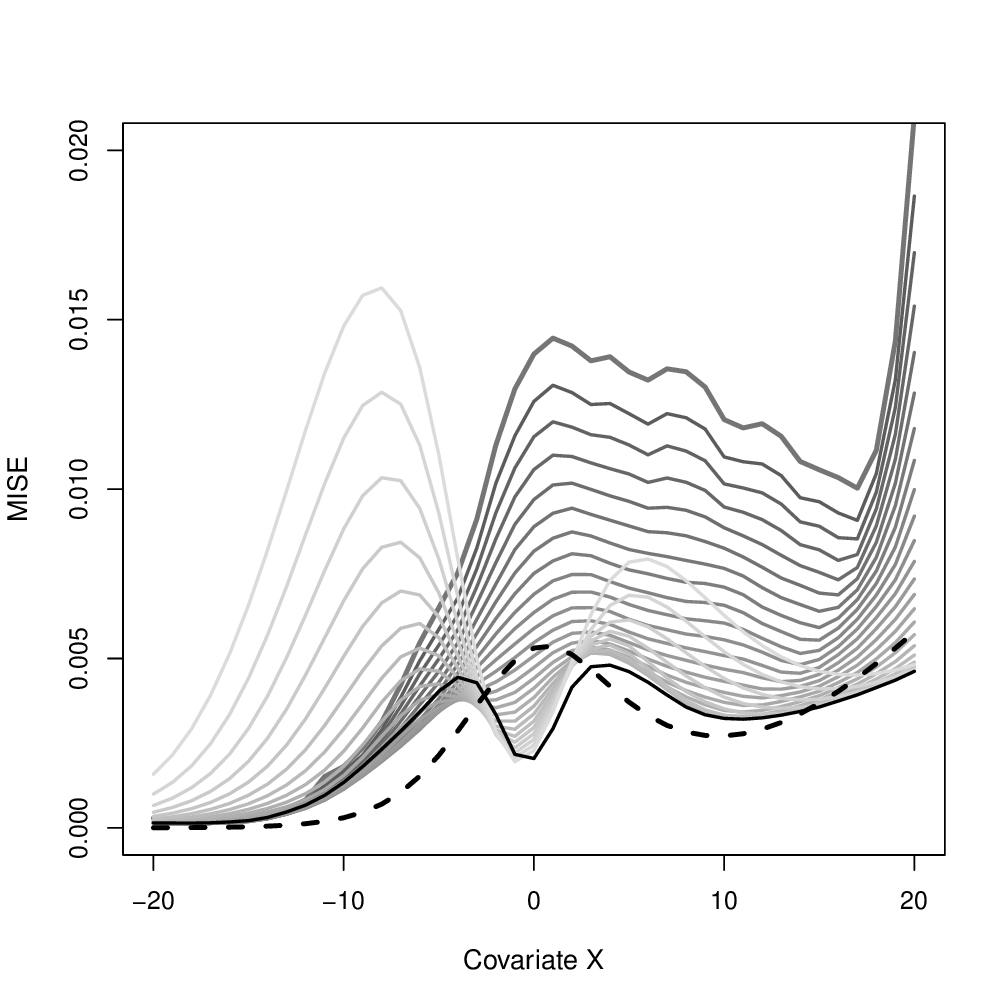}
	\end{subfigure}~
	\begin{subfigure}[b]{.5\textwidth}
	\includegraphics[width=\textwidth]{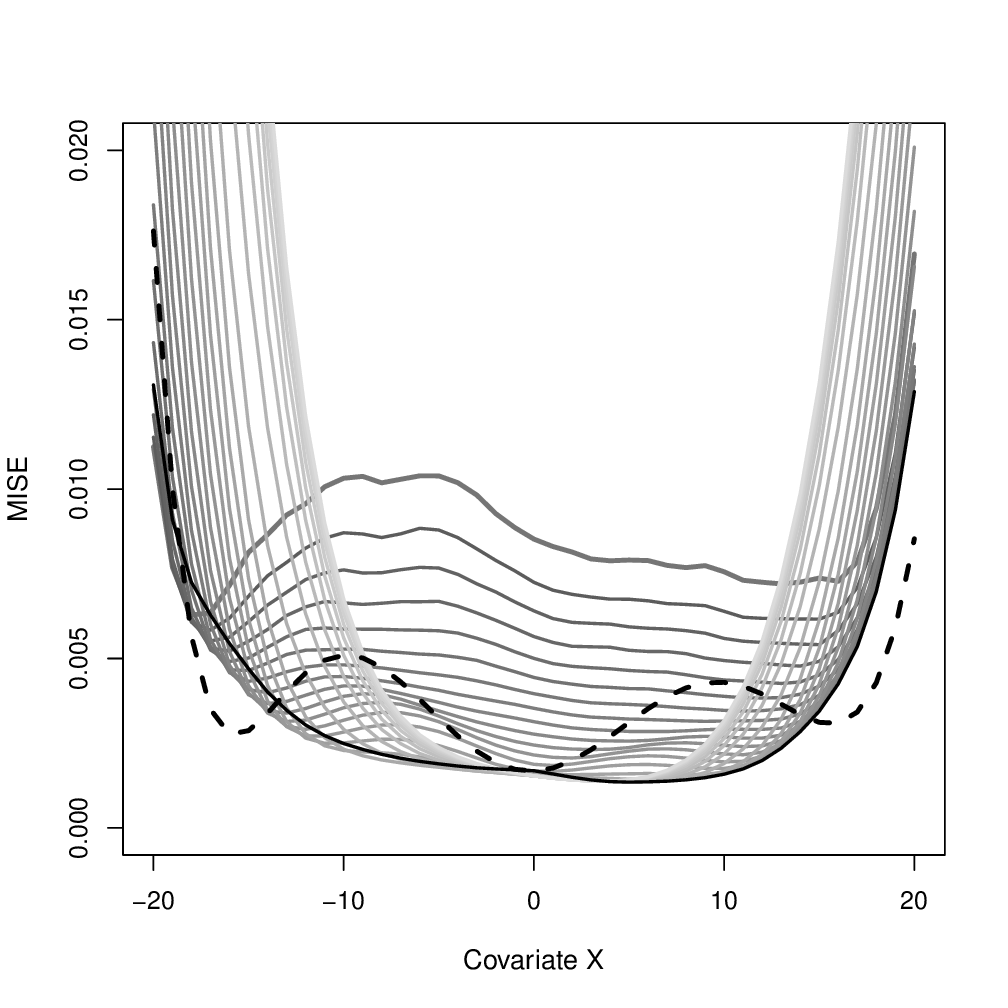}
\end{subfigure}
\caption{MISE of the proposed estimator $\widehat S_h^c(t| x)$ for a selection of $25$  bandwidths from the lowest (darkest grey line) to the highest (lightest grey line) in Scenario 1 (left) and Scenario 2 (right). Also shown are the MISE of  $\widehat S_h(t| x)$ computed with the optimal bandwidth (solid black line), and of the estimator by \cite{bernhardt2016flexible} (dashed black line).} \label{Fig_MSE}
\end{figure}

Figure \ref{Fig_MSE}  shows  the MISE curves of the three estimators.   In Scenario 1, as expected, the semiparametric estimator behaves well. Nevertheless, both $\widehat S_h^c(t| x)$ and  $\widehat S_h(t| x)$ are quite competitive for suitable values of the bandwidth, even beating the semiparametric estimator for some values of $X$  close to $0$ and $20$. In Scenario 2, both nonparametric estimators outperform the semiparametric estimator. Taking into account the known cure status gives either similar or better results than ignoring it for most values of $X$, especially in Scenario 2 (see Figure \ref{Fig_MSE}). In Table \ref{tab1:MISE}, the performance of the estimators is compared in terms of the integrated squared bias, integrated variance and MISE for the covariate values $x=-10,0$ and $10$. 
  In both scenarios, at $x=-10$,  the proposed estimator has smaller integrated squared bias and  variance than  Beran's estimator.  On the contrary, for $x=10$, the integrated squared bias and variance of  Beran's estimator is smaller compared to $\widehat S_h^c(t| x)$ estimator. As expected, the  integrated squared bias and   variance estimates for the semiparametric estimator are larger in Scenario 2. 

%============================
%\subsection{Bootstrap bandwidth selection results}
%============================

\begin{figure}[htbp]
	\begin{subfigure}[a]{.45\textwidth}
		\includegraphics[width=\textwidth]{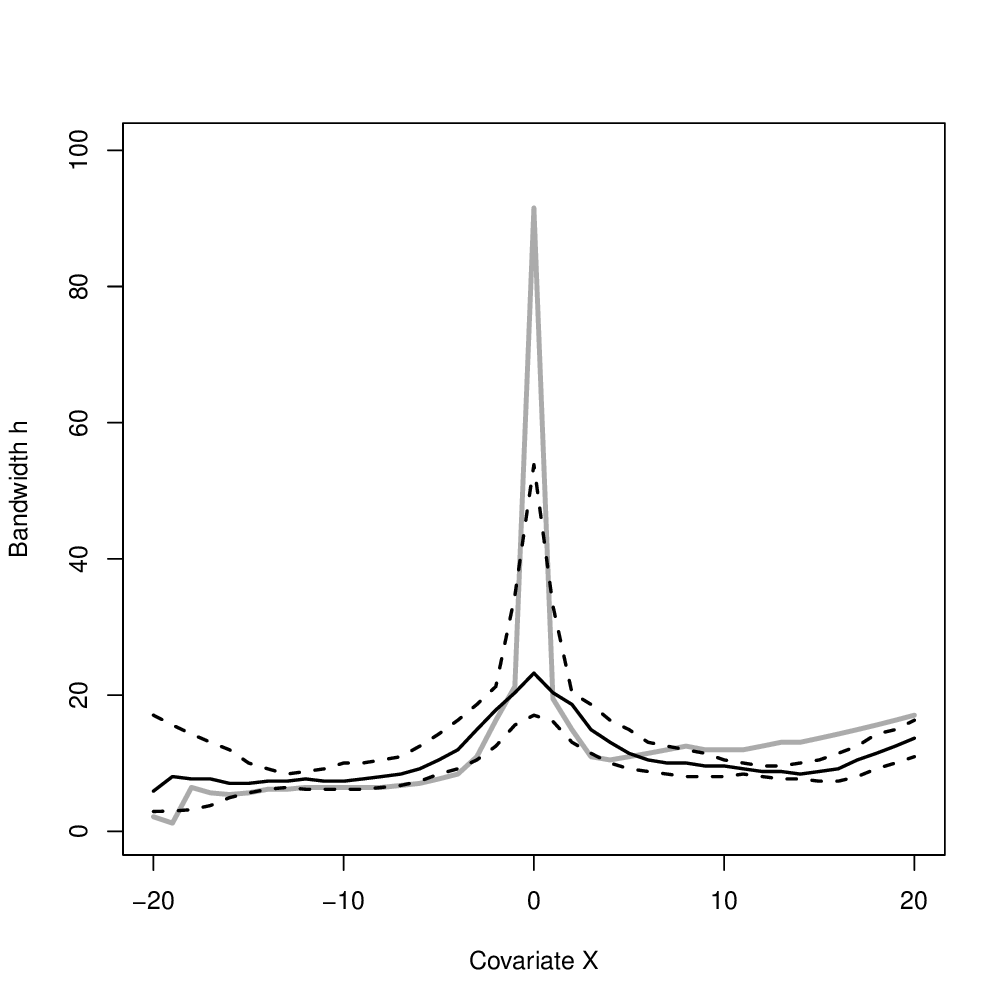}
	\end{subfigure}
	\begin{subfigure}[a]{.45\textwidth}
		\includegraphics[width=\textwidth]{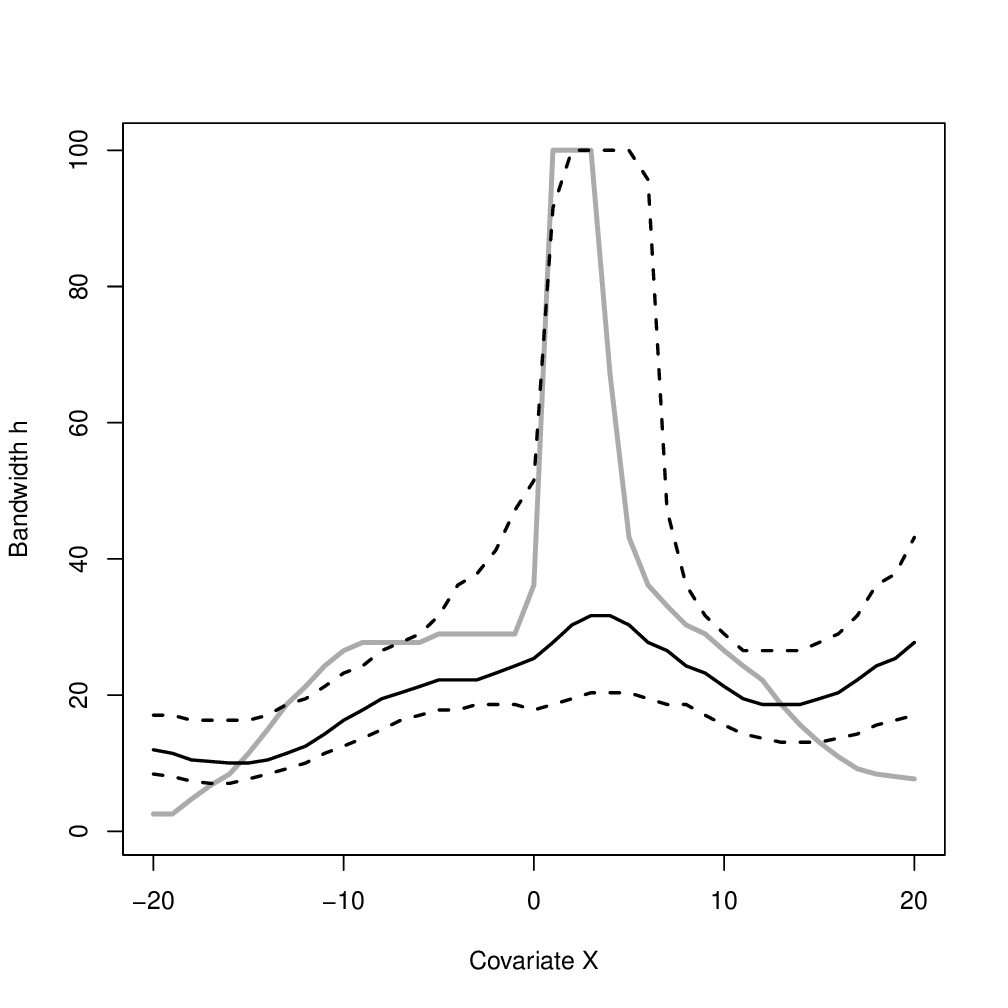}
	\end{subfigure}
	\caption{Median (solid black line) and first and third quartiles (dashed lines)  of the bootstrap bandwidths for $\widehat S_h^c(t| x)$ in Scenario 1 (left) and Scenario 2 (right). The optimal bandwidth (solid grey line) is displayed as reference. } \label{Fig_h_boot}
\end{figure}
\begin{table}[htbp]
	\centering
	\caption{Integrated squared bias ($\text{Ibias}{^2}$), integrated variance ($\text{Ivar}$) and MISE of the proposed estimator, $\widehat S_h^c(t|x)$, Beran's estimator, $\widehat S_h(t| x)$, (both computed with the optimal bandwidth) and the semiparametric estimator by \cite{bernhardt2016flexible}.}
	\resizebox{\textwidth}{!}{
		\begin{tabular}{crccccccccccccc}
			\toprule
		&	      & \multicolumn{4}{c}{\textbf{Proposed}}     &   & \multicolumn{4}{c}{\textbf{Beran}} &  &      \multicolumn{3}{c}{\textbf{Semiparametric}} \\
			\cmidrule{3-6}\cmidrule{8-11}\cmidrule{13-15}    \multirow{2}{*}{\textbf{Scenario}} & \multirow{2}{*}{$x$} & \multirow{1}{*}{$h$}& \multicolumn{1}{c}{$\text{Ibias}{^2}$  } & \multicolumn{1}{c}{$\text{Ivar}$} & \multicolumn{1}{c}{MISE} & &  \multirow{1}{*}{$h$}    & \multicolumn{1}{c}{$\text{Ibias}{^2}$} & \multicolumn{1}{c}{$\text{Ivar}$} & \multicolumn{1}{c}{MISE} &       & \multicolumn{1}{c}{$\text{Ibias}{^2}$} & \multicolumn{1}{c}{$\text{Ivar}$} & \multicolumn{1}{c}{MISE} \\
			&&&$\times10^{3}$&$\times10^{3}$&$\times10^{3}$&&&$\times10^{3}$&$\times10^{3}$&$\times10^{3}$&&$\times10^{3}$&$\times10^{3}$&$\times10^{3}$\\
			\midrule
		\multirow{3}{*}{$1$}	 &    $-10$ & $6.582$ &   $0.119$     &  $1.022$    &  $1.141$     &       & $6.334$  &  $0.163$   &  $1.177$     &    $1.340$   &       &  $0.002$     &  $0.299$     & $0.301$ \\
			&   $0$   & $20.000$  &   $0.371$    &   $1.834$    &    $2.205$   &       & $20.000$ &  $0.119$    &  $1.927$    &    $2.046$    &       &  $0.035$     &  $5.280$     & $5.315$ \\
			&   $10$   &  $12.152$ &   $0.375$    &   $2.902$    &    $3.277$   &       &  $12.387$&  $0.355$   &    $2.885$   &   $3.240$    &       &   $0.365$    &  $2.340$     & $2.705$ \\
			\midrule
			\multirow{3}{*}{$2$} &    $-10$  & $25.874 $ &   $0.076$    &    $2.206$   &    $2.282$   &       &  $23.492$&   $0.065$  &   $2.501$    &    $2.566$   &       &     $3.962$  &   $1.122$    & $5.084$ \\
			&    $0$   &$36.867$ &    $0.058$   &   $1.517$    &    $1.575$   & & $30.392$ & $0.151$   &   $1.632$    &     $1.783$  &               &  $0.034$    &    $1.641$   & $1.675$ \\
			&   $10$   &  $26.721$&  $0.103$     &    $1.474$   &    $1.577$   &   & $28.497$   &   $0.058$   &   $1.492$    &  $1.550$     &       &  $1.026$    &    $3.282$   & $4.308$ \\
			\bottomrule
	\end{tabular}}
	\label{tab1:MISE}%
\end{table}%

\begin{figure}
	\begin{subfigure}[c]{.45\textwidth}
		\includegraphics[width=\textwidth]{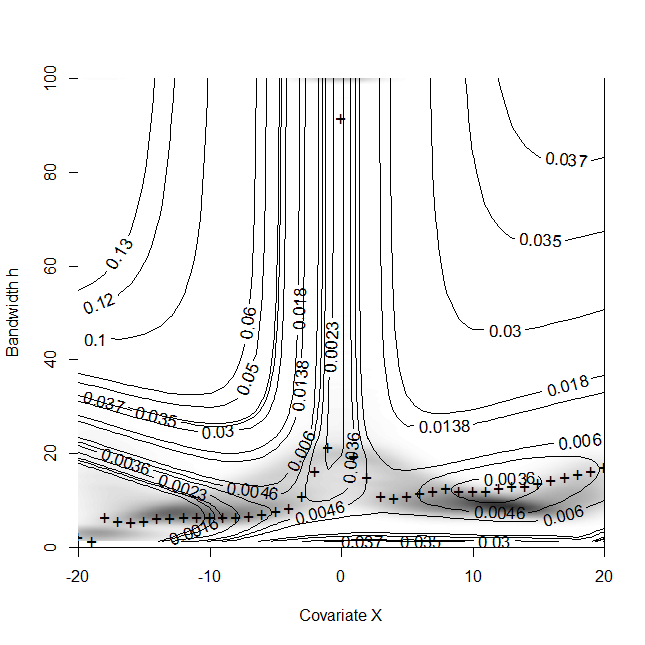}
		
	\end{subfigure}
	\begin{subfigure}[d]{.45\textwidth}
		\includegraphics[width=\textwidth]{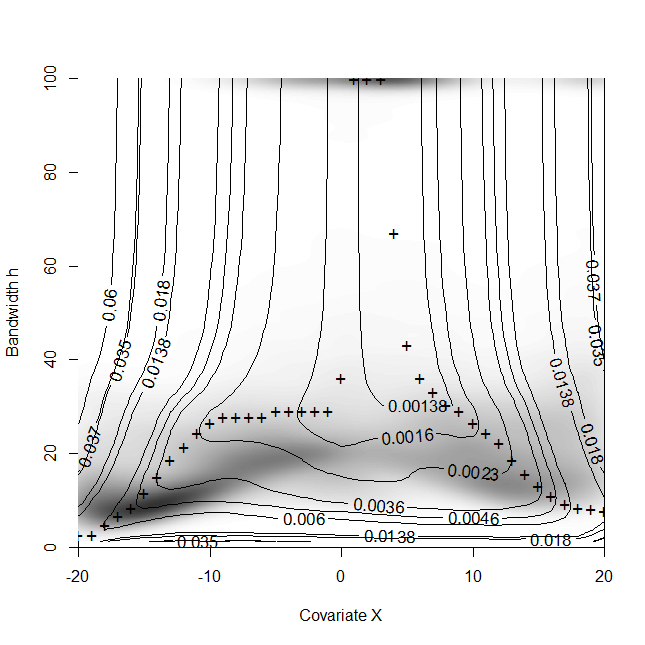}
		
	\end{subfigure}
\caption{Contour plots of the MISE of $\widehat S_h^c(t| x)$ as a function of the bandwidth $h$ and the covariate value  $x$ in Scenario 1 (left) and Scenario 2 (right).  For each value of covariate, the optimal bandwidth is marked with a cross. The density of the bootstrap bandwidths is shown in grey scale.}
\label{fig:cont}
\end{figure}

\begin{figure}
	\begin{subfigure}[c]{.45\textwidth}
		\includegraphics[width=\textwidth]{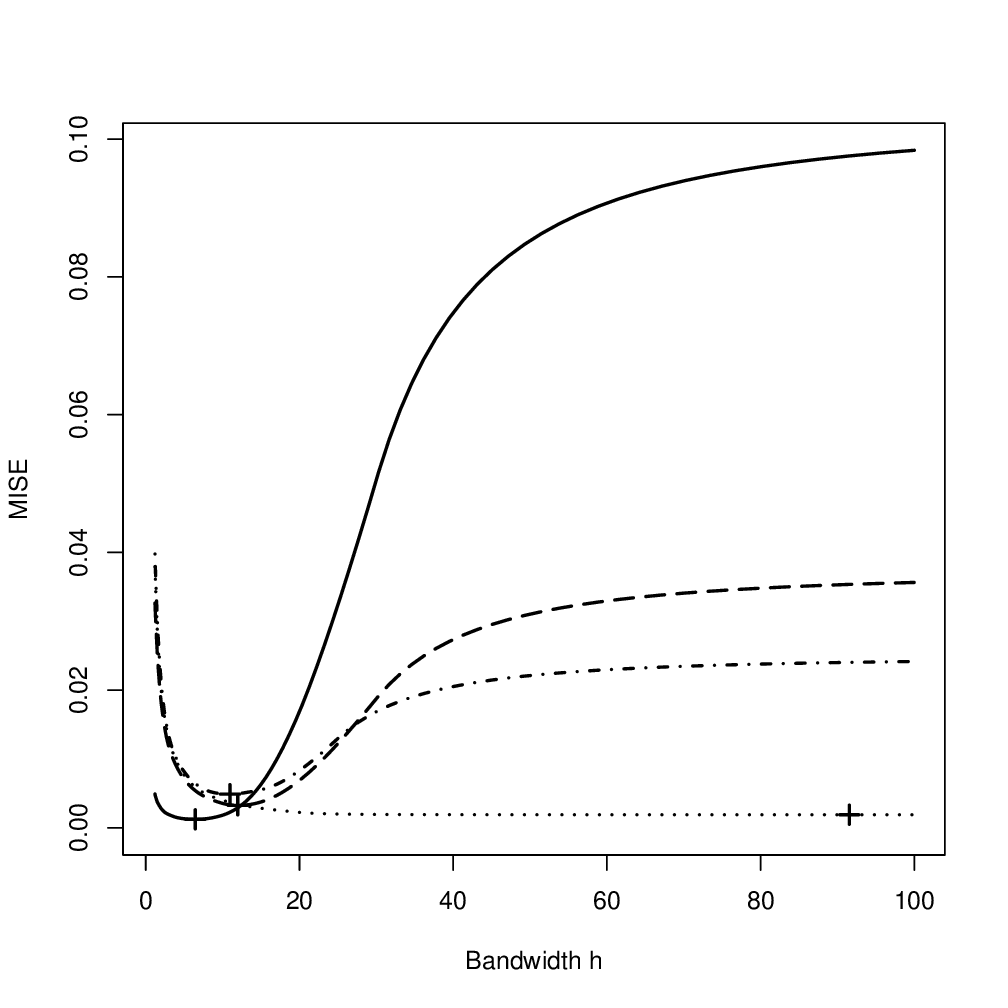}
		
	\end{subfigure}
	\begin{subfigure}[d]{.45\textwidth}
		\includegraphics[width=\textwidth]{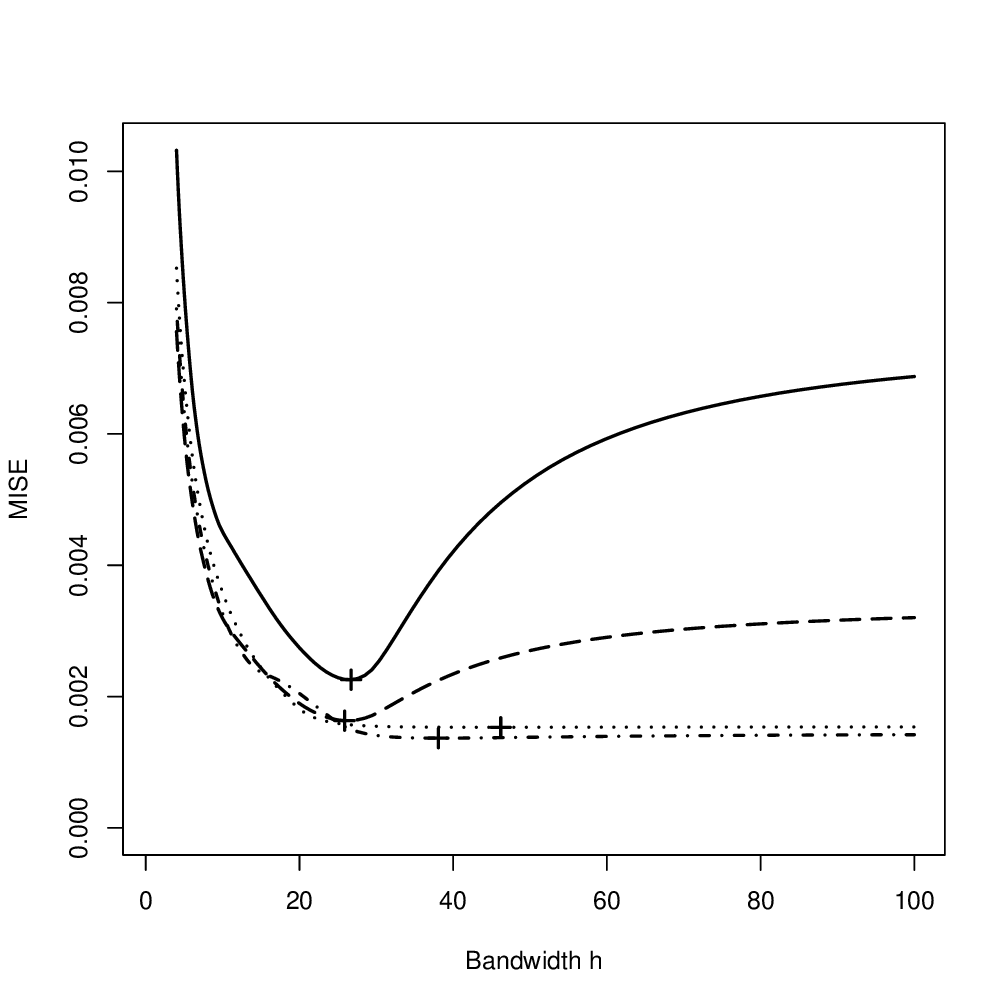}
		
	\end{subfigure}
	\caption{MISE of $\widehat S_h^c(t| x)$ as a function of the bandwidth $h$ for four different values of the covariate $x=-10$ (solid line), $x=0$ (dotted), $x=5$ (dot-dashed) and $x=10$ (long dash) in Scenario 1 (left) and Scenario 2 (right).  For each value of covariate, the  optimal bandwidth where the minimum MISE is reached is marked with a cross.  }
	\label{fig:miseh}
\end{figure}
The performance of the bootstrap bandwidth selector was assessed using $B=1000$ resamples. 
Figure \ref{Fig_h_boot} displays  the quartiles of the selected bootstrap bandwidths together with the optimal bandwidth.  Corresponding contour plots  in Figure \ref{fig:cont} show the density of the bootstrap bandwidths and the MISE of $\widehat S_h^c(t| x)$ as a function of the bandwidth $h$ and the covariate value $x$.   Figure \ref{fig:miseh} shows the MISE of $\widehat S_h^c(t| x)$ as a function of the bandwidth $h$, for four  values of the covariate.  Figure \ref{Fig_h_boot} and Figure \ref{fig:cont} illustrate that the bootstrap bandwidth approximates quite well the optimal bandwidth. Note that in Figure \ref{fig:cont} vertical contour lines indicate that, given $x$, the MISE of $\widehat S_h^c(t| x)$ tends to be constant as a function of $h$. Therefore, different bandwidths would yield approximately the same MISE. In those cases, the bootstrap bandwidth being far from the optimal bandwidth does not imply a loss of efficiency.  Similar results are observed 
 in Figure \ref{fig:miseh}. For example, let us consider $x=0$ in Scenario 2, we see that the MISE initially decreases as the bandwidth increases, although afterwards it becomes constant. 
%============================
\section{Application to  real data} \label{sec5:analysis}
%============================
To illustrate the practical performance of $\widehat S_h^c(t| x)$ we considered a dataset of $233$ patients of sarcoma cancer  aged $20 - 90$ years old from the University Hospital of Santiago de Compostela, Spain (CHUS).  Sarcoma is a rare type of cancer that represents $1\%$ of all adult solid malignancies. If a tumor can be surgically removed to render the patient with sarcoma free of detectable disease, $ 5$ years is the survival time at which  sarcoma oncologists assume long‐term remissions \citep{choy2014sarcoma}. Overall, $59$ patients died from sarcoma, and the remaining $174$  patients  were censored. Among  censored patients, $18$ patients were tumor free for more than five years. Hence, they were assumed to be long-term survivors. The aim was to estimate the survival time of the patients until death from sarcoma as a function of covariates such as the age at diagnosis, sex, tumor site, cancer spread (metastasis) and the margin status. The variables selected for estimating the survival probabilities were previously reported to be related to  long-term sarcoma survival \citep[][among others]{carbonnaux2019very,daigeler2014long}. 

Table \ref{tab2:cov} shows the descriptive demographic and clinical characteristics of  sarcoma patients  by age, sex and relevant clinical factors. Of the  $233$ patients, $100 ~(42.9\%)$  were males. Tumor site was categorized as retroperitoneal, extremities, or other. Most tumors were found in the retroperitoneum $(37\%)$ and in the extremities  $(30\%)$, with other areas of the body accounting for about $33\%$. Fifty-five $(32.9\%)$ patients were diagnosed of metastatic sarcoma. 

\begin{figure}[h]
	\begin{subfigure}{\textwidth}
		\includegraphics[width=.335\textwidth]{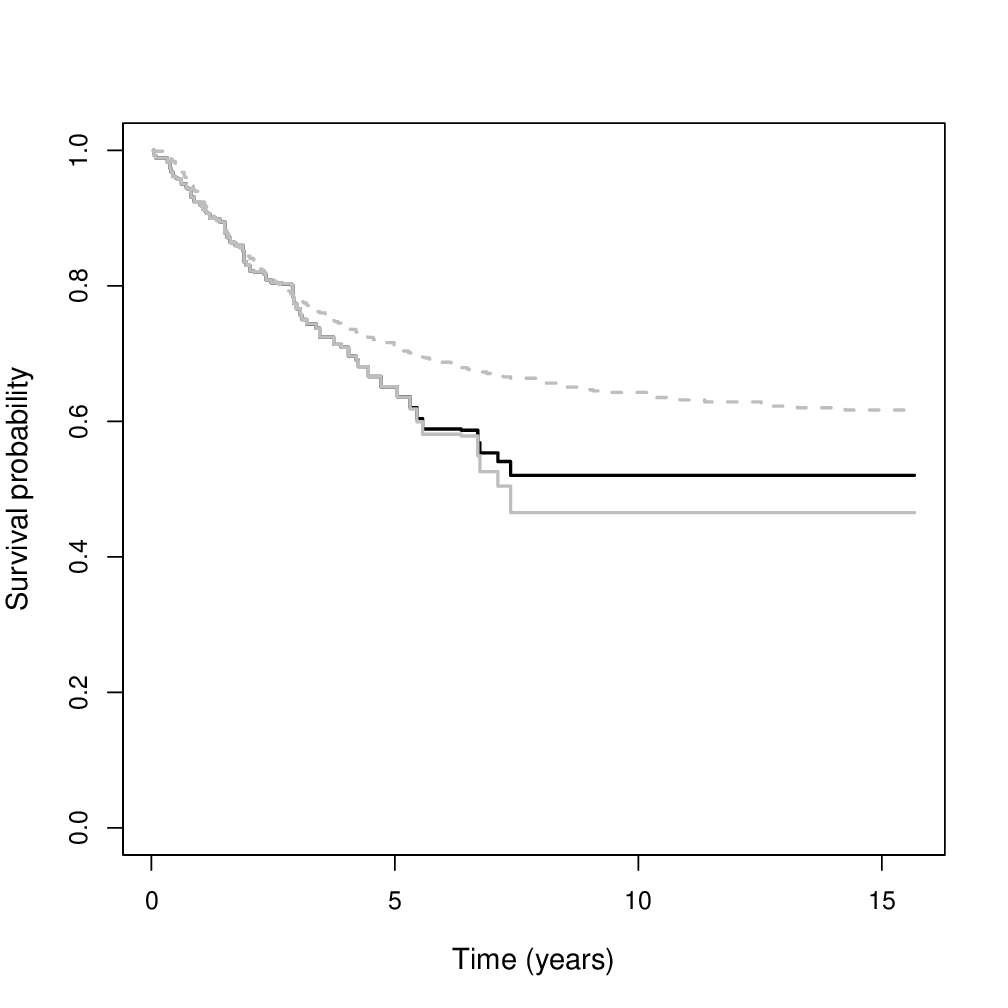}
		\includegraphics[width=.335\textwidth]{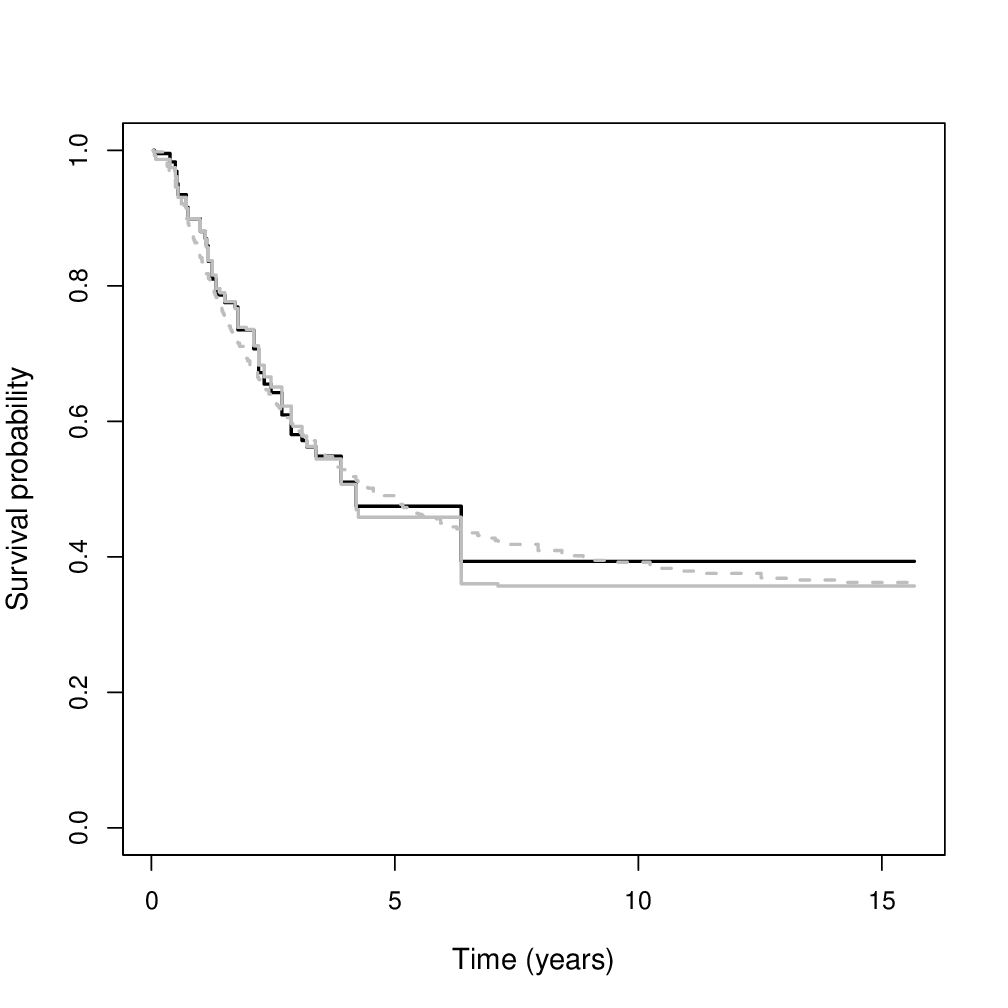}
		\includegraphics[width=.335\textwidth]{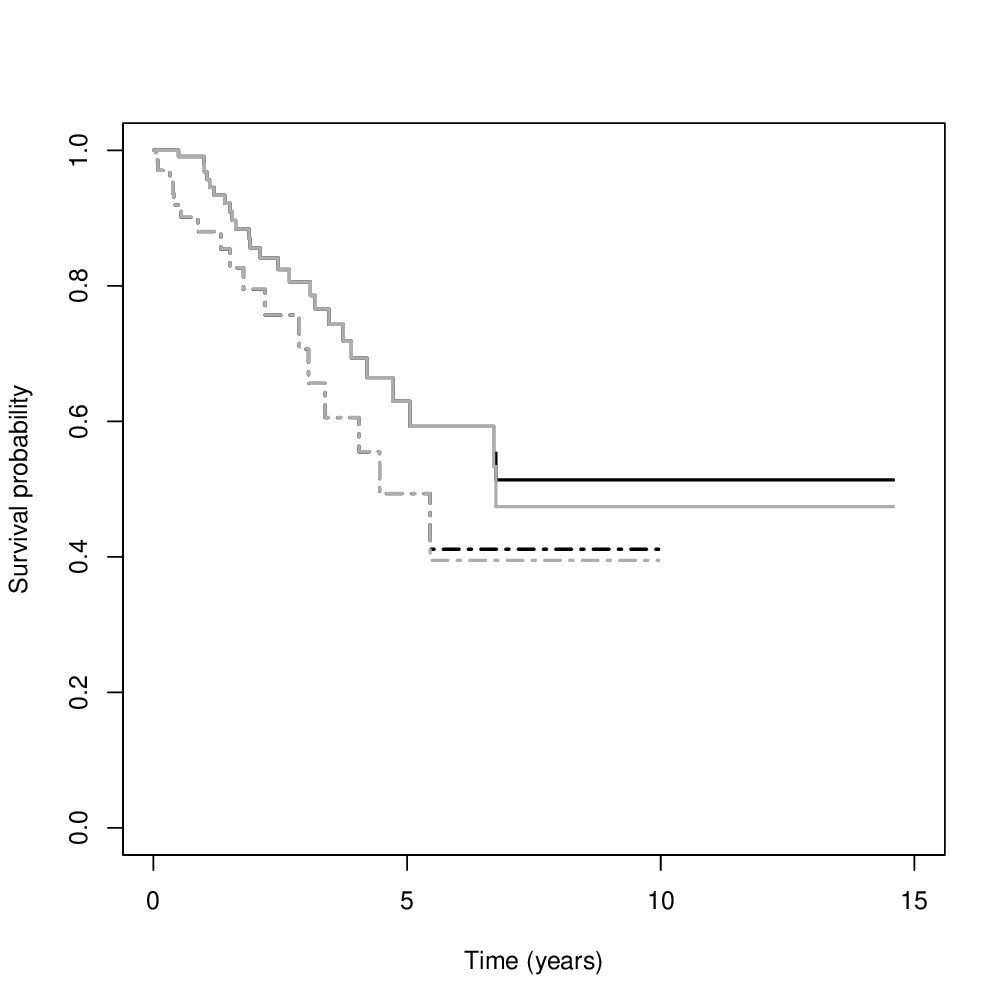}
	\end{subfigure}
	\caption{ Survival estimates for sarcoma  patients aged 40 (left) and 90 (center) years are obtained with the proposed estimator $\widehat{S}^c_h(t| x)$  (solid black  line) and  Beran's estimator $\widehat{S}_h(t| x)$ (solid grey  line),  both computed with the corresponding bootstrap bandwidth, and the semiparametric  estimator of \cite{bernhardt2016flexible} (dashed grey line). The right figure shows survival estimates stratified by the margin status, negative margin (solid lines) versus positive  margin (dashed lines). These estimates are computed  using the proposed estimator $\widehat{S}_n^c(t)$ (black lines) and the Kaplan-Meier estimator (grey lines).  } \label{Fig_sarcoma}
\end{figure}

Figure \ref{Fig_sarcoma} compares the results obtained by using the proposed estimator $\widehat S_h^c(t| x)$, which takes into account the 18 long-term survivors, with Beran's estimator $\widehat S_h(t| x)$, which  ignores individuals known to be cured and treats them as simply censored observations. Both estimators were computed with the corresponding bootstrap bandwidth. The semiparametric estimator of \cite{bernhardt2016flexible} was also considered as reference. All estimators show that the survival curve decreases when age increases from $40$ to $90$ years.  We find the largest differences between the proposed estimator and Beran’s estimator at the right tail of the distribution, where the survival curve for $\widehat{S}^c_h(t| x)$ is slightly higher. Since the cure probability can be obtained as the limit of $S(t| x)$ when $t\rightarrow \infty $ using the proposed
estimator of the survival curve will yield in higher estimates of the probability of cure. 

On the other hand,  the  survival curve estimated by the semiparametric estimator of \cite{bernhardt2016flexible} tends to decrease much slower than those obtained with the nonparametric estimators $\widehat{S}^c_h(t| x)$  and $\widehat{S}_h(t| x)$, suggesting that further testing is required to provide evidence that  assumptions in the semiparametric model  are fulfilled.

	 Figure \ref{Fig_sarcoma}  on the right shows the survival curves of sarcoma patients stratified  by the margin status.  In this case,  the proposed estimator in an unconditional setting $\widehat{S}_n^c(t)$ is applied and  the  \cite{kaplan1958nonparametric} estimator is considered as  reference.   The survival curves tend to decrease with time  in both  subgroups. The positive margin survival curve  decreases  slightly faster than the negative survival curve. In addition, the distinction between $\widehat{S}_n^c(t)$ and the Kaplan-Meier estimator is found at the right tail of the distribution with the survival curves estimated by $\widehat{S}^c_n(t)$ being slightly higher than the Kaplan-Meir curves.  For example, the survival probability, at the tail of the distribution,  for patients with negative margins   is around $0.51$ when estimated by $\widehat{S}_n^c(t)$, while it  is around $0.47$ when estimated by the Kaplan-Meier estimator. 
Again, the estimated probability of cure is slightly higher when the survival curve is fitted taking into account the known cured subjects.
\begin{table}[htbp]
	\centering
	\caption{Descriptive demographic and clinical characteristics of sarcoma patients stratified by age, sex, location of the sarcoma, metastatic and the margin status. In addition, the total number of patients for each subgroup ($n$), the number of patients died of sarcoma (death), those who were known to be cured (cured) and those with unknown cure status (unknown) are  given.}
	\begin{tabular}{lrccccc}
		\toprule
		\multicolumn{1}{l}{\multirow{2}[4]{*}{\textbf{Characteristics}}} & \multicolumn{1}{c}{\multirow{2}[4]{*}{\textbf{$n$ ($\%$)}}} & \multirow{2}[4]{*}{} & \multicolumn{1}{c}{\multirow{2}[4]{*}{\textbf{Death}}} & \multirow{2}[4]{*}{} & \multicolumn{2}{c}{\textbf{Censored}} \\
		\cmidrule{6-7}          &       &       &       &       & \multicolumn{1}{l}{\textbf{Cured}} & \multicolumn{1}{l}{\textbf{Unknown}} \\
		\midrule
	
	\textbf{Age}$^\dagger$ &       &  &  &  &       &    \\
\quad	$<60$   &  $105 ~(45.3\%)$     &       &   $25$    &       &  $9$     &   $71$    \\
\quad	$\ge 60$   &   $127~ (54.7\%)$    &       &    $33$   &       &    $9$   &    $85$   \\
	\textbf{Sex} &       &       &       &       &       &        \\
\quad	Male  &     $100 ~(42.9\%)$   &      &    $25$    &       &    $7$   &   $68$     \\
\quad	Female &     $133~ (57.1\%)$  &       &  $34$     &       &   $11$    &  $88$      \\  
	\textbf{Tumor site$^\dagger$} &       &       &       &       &       &       \\
\quad	Retroperitoneal &     $86 ~(37.2\%)$  &       &   $28$    &       &  $4$     &   $54$     \\
\quad	Extremities &  $70 ~(30.3\%)$     &       & $14$      &       &  $5$     &   $51$      \\
\quad	Other sites&      $75 ~(32.5\%)$ &       &   $16$    &       &    $9$   &  $50$      \\
	\textbf{Metastatic$^\dagger$} &       &       &       &       &       &        \\
\quad	No &   $112 ~(67.1\%)$    &       &    $11$   &       &   $9$    &    $92$    \\
\quad	Yes &    $55~ (32.9\%)$   &       &    $32$   &       &    $3$   &   $20$     \\
	\textbf{Margin status$^\dagger$} &       &       &       &       &       &       \\
\quad	Negative &   $133~ (65.8\%)$    &       &  $25$     &       &   $12$    &   $95$      \\
\quad	Positive &     $69~ (34.2\%)$  &       &    $17$   &       &    $3$   &   $49$     \\	\bottomrule
	\multicolumn{7}{l}{$^\dagger$ Contains a  few missing data.} \\
	
	\end{tabular}%
	\label{tab2:cov}%
\end{table}%

\section{Discussion} \label{sec6:discussion}
The proposed nonparametric estimator of the survival function takes advantage of the additional cure status information that Beran's estimator ignores. As a further step,  it could be used to derive nonparametric estimators for the cure probability and the latency function. 

Thus far, the estimation procedure was discussed involving a single continuous covariate. It would be of interest to extend our estimator to the case of multiple covariates, with $\mathbf{X}$  a vector of mixed discrete, categorical and/or continuous variables.  One possibility is to consider product kernels   (\citeauthor{li2008nonparametric}, \citeyear{li2008nonparametric}). 
  Another possibility is  to use  dimension reduction techniques like a single-index model. Specifically, the idea is to apply the proposed estimator of the survival function with a new covariate given by an estimator of the index $\widetilde{\mathbf{X}}=\mathbf{\beta}^T\mathbf{X}$, with $\mathbf{\beta}$ a parameter vector of the same dimension of $\mathbf{X}$. Semiparametric index estimation of the conditional distribution in the presence of right censoring was considered recently by  \cite{li2018dimension}.

Although the proposed estimator utilizes the cure status  information and shows good results both theoretically and practically, it is not without limitations.  It is competitive over Beran's estimator in terms of the MISE, showing a general better behavior. But for some values of the covariate it does not result in an improvement but a slightly worse MISE performance. The clear gain in terms of the integrated variance could be cancelled out by the integrated squared bias, which depends on the cure probability, the conditional censoring distribution and the conditional probability of observed cured individuals. 
   For the semiparametric estimator by \cite{bernhardt2016flexible}, our numerical experience indicates that if the sample size is small (less than 100), it is challenging to obtain stable values for the  model parameters.

  The \texttt{R} package \texttt{npcure} by
 \cite{LopezCheda2020} provides the nonparametric  estimation and testing procedures in mixture cure models proposed by \citeauthor{lopez2017nonparametric} \citeyearpar{lopez2017nonparametric,lopez2017bnonparametric,lopez2020nonparametric}, including the Beran's estimator. The situation when cure status is partially known is not currently supported by the package but will be considered in future versions. Further, the estimator of the conditional  survival function introduced in this paper and subsequent estimators of the cure rate and latency functions will be incorporated in the upgraded package.
%=================================
\section*{Acknowledgement}
%=================================
We are grateful to Dr. \'Angel D\'iaz-Lagares, Head of Cancer Epigenomics Lab, Translational Medical Oncology Group (IDIS, CHUS) and funded by a contract ``Juan Rod\'es'' (JR17/00016) from ISCIII, for providing with the sarcoma dataset obtained from the public The Cancer Genome Atlas (TCGA) program. This work has been supported by MINECO grant MTM2017-82724-R, the Xunta de Galicia (Grupos de Referencia Competitiva ED431C-2016-015) and the Centro de Investigación de Galicia ``CITIC'', funded by Xunta de Galicia and the European Union (European Regional Development Fund-Galicia 2014-2020 Program), by grant ED431G 2019/01, all of them through the ERDF.
%=================================
\section*{Conflict of interest}
%=================================
The authors have declared no conflicts of interest.

%=================================
\section*{ORCID}
%=================================
Wende Clarence Safari \orcidicon{0000-0003-4639-7552} \href{https://orcid.org/0000-0003-4639-7552}{\textcolor{blue}{https://orcid.org/0000-0003-4639-7552}} \\ 
Ignacio L\'opez-de-Ullibarri \orcidicon{0000-0002-3438-6621} \href{https://orcid.org/0000-0002-3438-6621}{\textcolor{blue}{https://orcid.org/0000-0002-3438-6621}} \\ M. Amalia Jácome \orcidicon{0000-0001-7000-9623} \href{https://orcid.org/0000-0001-7000-9623}{\textcolor{blue}{https://orcid.org/0000-0001-7000-9623}}

%=================================	
\section*{SUPPORTING INFORMATION}\label{SM}
%=================================
Supplementary Material  includes the proofs of Theorem \ref{thm:lambda}--\ref{thm:FhatFweigtht} and Lemmas 1--5 which are needed for deducing the asymptotic  bias and variance of $\widehat{S}_h^c(t| x)$.  It also includes: the simulation results for  sample sizes $n=50$, $100$, $200$, and  proportions of observed cured data $\pi=0.2, 0.8$, $1$; and additional simulations to study the effect of the pilot bandwidth on the estimates. Additional supporting information including source code to reproduce the results can be found online in the Supporting Information section at the end of the article.
%=================================
\appendix
%=================================
 \noindent The following assumptions are made.
\begin{assumption}\label{ass:A1}\normalfont	\begin{itemize}
		\item [(i)]  Let $I=\left[x_1,x_2\right]$ be an interval contained in  the support of the density function of $X$, $m(x)$, such that
		$$0<\gamma=\inf_{x\in I_\varepsilon} m(x)<\sup_{x\in I_\varepsilon}m(x)=\Gamma<\infty$$ 	for some $I_\varepsilon=[x_1-\varepsilon,x_2+\varepsilon]$ with $\varepsilon>0$ and $0 < \varepsilon\Gamma<1$. And for all $x\in I$,  $Y$,  $C$ are conditionally independent at $X=x$.
		\item[(ii)] 	There exist	 $a,b\in \mathbb{R}$, with $a<b$ satisfying
		$J\left(t| x\right)\ge \theta > 0 \text{ \ for \ } (t,x)\in [a,b]\times I_\varepsilon. $
	\end{itemize}
\end{assumption}
\begin{assumption}
	\label{ass:A3} \normalfont The first derivative with respect to $x$ of function $m(x)$ exists and is continuous  in $x \in I_\varepsilon$, and the first derivatives with respect to $x$ of functions
	$H(t| x)$, $H^1(t| x)$ and $H^{11}(t| x)$  exist and are continuous and  bounded  in $(t,x) \in  [0, \infty) \times I_\varepsilon$.
\end{assumption}
\begin{assumption}
	\label{ass:A4} \normalfont
	The second derivative with respect to $x$ of function $m(x)$ exists and is continuous  in $x \in I_\varepsilon$, and the second derivatives with respect to $x$ of functions
	$H(t| x)$, $H^1(t| x)$ and $H^{11}(t| x)$  exist and are continuous and  bounded  in $(t,x) \in  [0, \infty) \times I_\varepsilon$.
\end{assumption}
\begin{assumption}
	\label{ass:A6} \normalfont
	The first  derivatives with respect to $t$ of  $H(t| x)$,  $H^{1}(t| x)$ and $H^{11}(t| x)$   exist and are continuous in $\left(
	t,x\right) \in \lbrack a,b] \times I_\varepsilon$.
\end{assumption}
\begin{assumption}
	\label{ass:A7} \normalfont
	The second  derivatives with respect to $t$ of  $H(t| x)$,  $H^{1}(t| x)$ and $H^{11}(t| x)$   exist and are continuous in $\left(t,x\right) \in \lbrack
	a,b]\times I_\varepsilon$.
\end{assumption}
\begin{assumption}
	\label{ass:A8}\normalfont
	The first  derivative with respect to $x$ and the second derivative with respect to $t$ of  $H(t| x)$,  $H^{1}(t| x)$ and $H^{11}(t| x)$   exist and are continuous in $(t,x)\in [a,b] \times I_\varepsilon$.
\end{assumption}
\begin{assumption}
	\label{ass:A9}\normalfont
	The (sub)densities corresponding to   the (sub)distribution functions $H(t| x)$,  $H^{1}(t| x)$ and $H^{11}(t| x)$ are bounded away from $0$ in $[a,b]\times I_\varepsilon$.
\end{assumption}
\begin{assumption}
	\label{ass:A10}\normalfont
	The kernel function $K(v)$ is a symmetrical density with zero mean,  vanishing outside $(-1, 1)$, and the total variation is less than  $\lambda < \infty$.
\end{assumption}

\noindent\textbf{Motivation of the proposed  estimators.} The cumulative hazard function $\Lambda(t|  x)$ can  be written as follows:
\begin{align}
\Lambda(t\mid x) =& \int_{0}^{t}\dfrac{dF\left(v\mid  x\right)}{1-F\left(v^-\mid  x\right)} 
=\int_{0}^{t}\dfrac{\left\{1-G(v^-\mid x)+G^{11}(v^-\mid x)\right\}dF(v\mid x)}{\left\{1-G(v^-\mid x)+G^{11}(v^-\mid x)\right\}\{1-F(v^-\mid x)\}}, \tag{A1}
\label{eq:Lambda}
\end{align}
where   $G^{11}(t| x)=P\left(C\le t,\xi=1,\nu=1| X=x\right)$ is the conditional  censoring subdistribution of the individuals observed to be cured.  The numerator in (\ref{eq:Lambda}) is $dH^1(t\mid x)$:
\begin{align}		
&\int_{0}^{t}\left\{1-G(v^-\mid x)+G^{11}(v^-\mid x)\right\}dF(v\mid x) \notag\\
&=\int_{0}^{t}P(C\ge v\mid X=x)dF(v\mid x)+\int_{0}^{t}P(C< v,\xi=1,\nu=1\mid X=x)dF(v\mid x)\notag\\
&=P(Y\le t,C\ge Y\mid X=x)  + P(Y\le t,C < Y, \xi=1,Y=\infty \mid X=x) \notag\\
&=P(T \le t,\delta=1\mid X=x)=H^1(t\mid x). \tag{A2}\label{App:dH1}
\end{align}
Similarly,  the denominator in (\ref{eq:Lambda}) is $J(t^-| x)$:
\begin{align}
&\left\{1-G(t^-\mid x)+G^{11}(t^-\mid x)\right\}\{1-F(t^-\mid x)\}\notag\\
&=\{P(C\ge t\mid X=x)+P(C<  t, \xi=1, \nu=1\mid X=x)\} P(Y\ge t\mid X=x)\notag\\
&=P(Y\ge t,C\ge t\mid X=x)+P(Y\ge t, C<  t,\xi=1, \nu=1\mid X=x)\notag\\
&=P(T\ge t\mid X=x)+P(T< t,\xi=1, \nu=1\mid X=x)\notag\\
&=1-H\left(t^-\mid x\right)+H^{11}\left(t^-\mid x\right)=J(t^-\mid x).\tag{A3}	\label{App:J}
\end{align}
Taking (\ref{App:dH1}) and (\ref{App:J}) into account, (\ref{eq:Lambda}) can be written as
\begin{align}
\Lambda(t\mid  x)=\int_{0}^{t}\dfrac{dH^1(v\mid  x)}{J(v^-\mid  x)}. \tag{A4}\label{eqn:lambda}
\end{align}
Consider the Nadaraya-Watson kernel estimates of $H^1(t| x)$ and $J(t^-| x)$:
\begin{align}
\widehat{H}_h^1(t\mid x)&=\sum_{i=1}^{n} B_{hi}(x) \textbf{1}\left(T_i\le t,\delta_{i}=1\right), \label{h1estimate}\tag{A5}\\	
\widehat{J}_h	(t^-\mid  x)&=\sum_{i=1}^{n}B_{hi}(x) \textbf{1}\left(T_i\ge t\right) +\sum_{i=1}^{n} B_{hi}(x) \textbf{1}\left(T_i<t,\xi_i\nu_i=1\right).\tag{A6}\label{Jestimate}
\end{align}
 The estimator of $\Lambda(t| x)$ when the cure status is partially known, $\widehat{\Lambda}^c_h(t| x)$, is obtained by plugging in (\ref{eqn:lambda}) the estimates (\ref{h1estimate}) and (\ref{Jestimate}). 
As for the estimator of the survival function, it can be shown that  $S(t| x)=\exp\left\{-\Lambda(t| x)\right\}$. By considering a Taylor’s expansion of the exponential function around 0 and evaluating it at each increment of $\widehat{\Lambda}_h^c(t| x)$, the estimator $\widehat{S}^c_h(t |  x)$ in (\ref{est_F}) is obtained.

\noindent\textbf{Proof of Proposition \ref{prop:property}}. The estimator $\widehat{S}_{h}^c(t| x)$ has  the following properties  

\begin{enumerate}
\item If there is no known cure status, $\widehat{S}_{h}^c(t| x)$ reduces to $\widehat{S}_{h}(t| x)$.
\begin{proof} It is straightforward since $\xi_i\nu_i=0$, $i=1,\ldots,n$. \end{proof}

\item In the specific case when some individuals are observed as cured when their survival time exceeds a known fixed cure threshold,  $\widehat{S}_{h}^c(t| x)$ reduces to $\widehat{S}_{h}(t| x)$.
\begin{proof} 
	Assume there exists a common specific
	known cure threshold $d_{i}=d$ for $i=1,\ldots,n$. This implies that in
	the 	ordered sample, $\left\{\left( X_{\left[ i\right] },T_{\left( i\right) },\delta _{%
		\left[ i\right] },\xi _{\left[ i\right] },\xi _{\left[ i\right] }\nu _{\left[
		i\right] }\right):i=1,\ldots,n\right\}$, the $n_{1}$ first observations correspond to
	individuals with $T_{\left( i\right) }<d$ either not cured or with unknown
	cure status ($\xi _{[i]}\nu _{[i]}=0$), and the remaining $m$ observations are
	cured individuals with $T_{\left( i\right) }\geq d$ and $\xi _{[i]}\nu _{[i]}=1$. Therefore,
$$
	\widehat{S}_{h}^c\left( t\mid x\right)  
	=\prod_{i=1}^n\left\{ 1-\frac{\delta _{\left[ i\right] }B_{h\left[ i%
			\right] }\left( x\right)\textbf{ 1}\left(T_{(i)}\le t\right)}{\sum_{j=i}^{n_{1}}B_{h\left[ j\right] }\left(
		x\right) +\sum_{j=n_{1}+1}^{n}B_{h\left[ j\right] }\left(
		x\right) }\right\}  =\prod_{i=1}^n\left\{ 1-\frac{\delta _{\left[ i\right] }B_{h\left[ i%
			\right] }\left( x\right) \textbf{1}\left(T_{(i)}\le t\right) }{\sum_{j=i}^{n}B_{h\left[ j\right] }\left(
		x\right) }\right\}=\widehat{S}_{h}(t\mid x)
	$$
This completes the proof.
 \end{proof}

\item When there is no censoring, the estimator $\widehat{S}_{h}^c(t| x)$ 
reduces to the kernel type estimator of the conditional survival function.
\begin{proof} Without censoring, $T_{i}=Y_{i},\delta _{i}=1$ and the cure
status is always observed $\xi _{i}=1$. In this situation, the $n=n_1+m$ observations can be ordered and split
 into the $n_1$ uncured individuals with finite lifetimes $Y_{i}$, and the $m$ cured individuals with lifetime $Y_{i}=\infty $.  Thus,
\begin{align*}
 \widehat{S}_{h}^c(t\mid x) =&\prod_{i=1}^{n}\left\{ 1-\frac{%
	B_{h[i]}\left( x\right) \textbf{1}\left( Y_{\left( i\right) }\leq t\right) }{%
	\sum_{j=i}^{n}B_{h[j]}\left( x\right)
          +\sum_{j=1}^{  i-1}B_{h[j]}\left(
	x\right) \textbf{1}\left( \nu _{\lbrack j]}=1\right) }\right\}   \\
=&\prod_{i=1}^{n}\left\{ 1-\frac{B_{h[i]}\left(
	x\right) \textbf{1}\left( Y_{\left( i\right) }\leq t\right) }{
	\sum_{j=i}^{n_{1}}B_{h[j]}\left( x\right)
	+\sum_{j=n_{1}+1}^{n}B_{h[j]}\left( x\right) }\right\} =\prod_{i:Y_{\left( i\right) }\leq t}\left\{ \frac{%
	\sum_{j=i+1}^{n}B_{h[j]}\left( x\right) }{\sum%
	_{j=i}^{n}B_{h[j]}\left( x\right) }\right\}.
\end{align*}
Note that the kernel estimator of the survival function  $ \widetilde{S}_{h}(t| x)=\sum_{i=1}^{n}B_{h[i]}\left( x\right) \textbf{1}(Y_{(i)}>t)
$  is a step function with jumps $B_{hi}\left( x\right) $ at the observations, $Y_{i}$. By defining $k=\max \{i:Y_{\left( i\right) }\leq t\}$ i.e., $Y_{\left( k\right)
}\leq t$ and $Y_{\left( k+1\right) }>t$, one can write

 \begin{align*}
\prod_{i:Y_{\left( i\right) }\leq t}\left\{ \frac{%
	\sum_{j=i+1}^{n}B_{h[j]}\left( x\right) }{\sum%
	_{j=i}^{n}B_{h[j]}\left( x\right) }\right\} &=\prod_{i:Y_{\left( i\right) }\leq t}\left\{
\frac{\widetilde{S}_{h}(Y_{(i)}\mid x)}{\widetilde{S}_{h}(Y_{(i-1)}\mid x)}\right\} 
=\frac{\widetilde{S}_{h}(Y_{\left( 1\right) }\mid x)}{1}\ \frac{
	\widetilde{S}_{h}(Y_{\left( 2\right) }|x)}{\widetilde{S}_{h}(Y_{\left(
		1\right) }\mid x)}\ \ldots \ \frac{\widetilde{S}%
	_{h}(Y_{\left( k\right) }\mid x)}{\widetilde{S}_{h}(Y_{\left( k-1\right) }\mid x)}
\\
&=\widetilde{S}_{h}(Y_{\left( k\right) }\mid x)=\sum_{i=1}^{n}B_{h[i]}\left( x\right)\textbf{1}(Y_{(i)}>t).
\end{align*}%
This completes the proof.
\end{proof}
\item In an unconditional setting  the proposed estimator is
\begin{align*}
	\widehat{S}_{n}^c\left( t\right)=\prod_{i=1}^{n}\left\{ 1-\frac{\delta
		_{\lbrack i]} \textbf{1}\left( T_{\left( i\right)
		}\leq t\right) }{n-i+1+  \sum_{j=1}^{i-1}\textbf{1}\left( \xi _{[j]}\nu _{[j]}=1\right)}\right\}. 
\end{align*}
\begin{proof}
	In unconditional setting the weights are $1/n$ for  $i=1,\ldots,n$. Thus, the proposed estimator becomes
	\begin{align*}
		\widehat{S}_{n}^c\left( t\right)=\prod_{i=1}^{n}\left\{ 1-\frac{\delta
			_{\lbrack i]}\frac{1}{n} \textbf{1}\left( T_{\left( i\right)
			}\leq t\right) }{\frac{1}{n}(n-i+1)+  \frac{1}{n}\sum_{j=1}^{i-1}\textbf{1}\left( \xi _{[j]}\nu _{[j]}=1\right)}\right\}. 
	\end{align*}
In the particular case where an individual
is known to be cured only if the observed time is greater than a known fixed time, say $d$, with $n = n_1 + m$ observations, when $m$ are identified as cured, the ordered observed lifetimes are $T_{(1)} \leq \ldots \leq T_{(n_1)}$ strictly lower than $d$, and the $m$ cured individuals with $T_{(i)}\ge d$.  Thus, $\widehat{S}_{h}^c(t| x)$ 
reduces to the generalized maximum likelihood estimator in \cite{laska1992nonparametric}:
\begin{align*}
 \widehat{S}_{n}^c(t)  &=\prod_{i=1}^{n}
\left\{ 1-\frac{\delta_{[i]}\frac{1}{n} \textbf{1}\left( T_{\left( i\right) }\leq t\right) }
{\frac{1}{n}(n_1 -i+1 )+  \frac{1}{n}m }\right\} = \prod_{i=1}^n
\left\{1-\frac{\delta_{[i]}\textbf{1}(T_{(i)}\leq t) }{n -i+1 }\right\}. \\
\end{align*}
This completes the proof.
\end{proof}
\end{enumerate}

\noindent\textbf{Proof of Proposition \ref{prop:npmle}.} The proof follows the argument in Theorem 2 in \cite{lopez2017nonparametric} and Theorem 1 in \cite{laska1992nonparametric}. To derive the
expression of the local likelihood of the mixture cure model, we consider the three
potential cases for the $i$th observation:

\begin{description}
\item[Case 1: $\delta_i=1$. ]The event is observed and the individual is not cured. We observe  $Y_{i}=t_{i},%
\protect\nu _{i}=0$, with probability:
\begin{align*}
P\left( Y_{i}=t_{i},C_{i}>t_{i},\nu _{i}=0\mid X=x\right)  
=&P\left( C_{i}>t_{i}\mid Y_{i}=t_{i},\nu _{i}=0,X=x\right)  \\
&\times P\left( Y_{i}=t_{i}\mid \nu _{i}=0,X=x\right) P\left( \nu
_{i}=0\mid X=x\right)  \\
=&S_{C\mid Y,X,\nu =0}\left( t_{i}\mid x\right)  \left\{
S_{0}(t_{i}^{-}\mid x)-S_{0}(t_{i}\mid x)\right\}  p\left( x\right),
\end{align*}
where $S_{C\mid Y,X,\nu =0}(t\mid x)$ is the  conditional survival function for the censoring variable $C$ for uncured individuals.

\item[Case 2: $ (\protect\delta _{i}=0,\protect\xi _{i}\protect\nu %
_{i}=0 )$. ] The individual is censored and the cure status is unknown. We observe $C_{i}=t_{i}$, and $
\protect\nu _{i}$ is unknown, with probability:
\begin{align*}
P\left( Y_{i}>t_{i},C_{i}=t_{i}\mid X=x\right)  
=&P\left( Y_{i}>t_{i},C_{i}=t_{i},\nu _{i}=1\mid X=x\right)P(\nu_i=1\mid X=x)\\ &+P\left(
Y_{i}>t_{i},C_{i}=t_{i},\nu _{i}=0\mid X=x\right)P(\nu_i=0\mid X=x)  \\
=&f_{C\mid Y,X,\nu =1}\left( t_{i}\mid x\right)  \left\{ 1-p\left(
x\right) \right\} +f_{C\mid Y,X,\nu =0}\left( t_{i}\mid x\right)\times S_{0}\left(
t_{i}\mid x\right) p\left( x\right), 
\end{align*}
where $f_{C\mid Y,X,\nu =1}(t\mid x)$ and $f_{C\mid Y,X,\nu =0}(t\mid x)$ are the conditional  density functions for the censoring variable $C$ of the cured and uncured individuals, respectively.

\item[Case 3: $\left( \delta _{i}=0,\xi _{i}\protect\nu %
_{i}=1\right) $. ] The individual is censored and known to be cured. We observe $C_{i}=t_{i}, \protect\nu _{i}=1$, with probability
\begin{align*}
P\left( Y_{i}>t_{i},C_{i}=t_{i},\nu _{i}=1\mid X=x\right)  
=&P\left( C_{i}=t_{i}\mid Y_{i}>t_{i},\nu _{i}=1,X=x\right) \\&\times P\left(
Y_{i}>t_{i}\mid \nu _{i}=1,X=x\right)  P\left( \nu _{i}=1\mid X=x\right)  \\
=&f_{C\mid Y,X,\nu =1}\left( t_{i}\mid x\right)  \left\{1-p\left(
x\right) \right\}.
\end{align*}
\end{description}
In the absence of specification of the distribution of $X$, the terms in the log-likelihood are weighted with the kernel weights $B_{h[i]}(x)$. Then, the local likelihood of the data is 
\begin{align*}
L\left( X,T,\delta,\xi,\nu \right)  
=&\prod\limits_{i=1}^{n}\left[ S_{C\mid Y,X,\nu =0}\left( T_{\left(
	i\right) }\mid x\right) \left\{ S_{0}(T_{\left( i\right) }^{-}\mid x)-S_{0}(T_{\left(
	i\right) }\mid x)\right\} p\left( x\right) \right] ^{B_{h\left[ i\right] }\left(
	x\right) \bm{1}\left( \delta _{\left[ i\right] }=1\right) }  \\
&\times \left[ f_{C\mid Y,X,\nu =1}\left( T_{\left( i\right) }\mid x\right) \left\{
1-p\left( x\right) \right\} +f_{C\mid Y,X,\nu =0}\left( T_{\left( i\right)
}\mid x\right) S_{0}\left( T_{\left( i\right) }\mid x\right) p\left( x\right)
\right] ^{B_{h\left[ i\right] }\left( x\right) \bm{1}\left( \delta _{\left[ i%
		\right] }=0,\xi _{\left[ i\right] }\nu _{\left[ i\right] }=0\right) } \\
& \times\left[ f_{C\mid Y,X,\nu =1}\left( T_{\left( i\right) }\mid x\right)
\left\{ 1-p\left( x\right) \right\}\right] ^{B_{h\left[ i\right] }\left(
	x\right) \bm{1}\left( \delta _{\left[ i\right] }=0,\xi _{[i]}\nu _{[i]}=1\right)}.
\end{align*}
If the distribution of the censoring variable $C$ is conditionally independent of $Y$ and the cure status $\nu $ given the covariate $X$, then 
\begin{align}
L\left( X,T,\delta,\xi,\nu \right)  
=&\prod\limits_{i=1}^{n} \left[q_i(x) p\left( x\right) \right] ^{B_{h%
		\left[ i\right] }\left( x\right) \textbf{1}\left( \delta _{\left[ i\right] }=1\right)
} \notag \times \left\{ 1-p\left( x\right)  +S_{0}\left( T_{\left(
	i\right) }\mid x\right) p\left( x\right) \right\} ^{B_{h\left[ i\right] }\left(
	x\right) \textbf{1}\left( \delta _{\left[ i\right] }=0,\xi _{\left[ i\right] }\nu _{%
		\left[ i\right] }=0\right) }\notag \\
& \times \left\{ 1-p\left( x\right) \right\} ^{B_{h\left[ i\right]
	}\left( x\right) \textbf{1}\left( \delta _{\left[ i\right] }=0,\xi _{\lbrack i]}\nu
	_{\lbrack i]}=1\right) }\notag \times\left( 1-\sum_{j=1}^{i-1}g_{j}\left(
x\right) \right) ^{B_{h\left[ i\right] }\left( x\right) \textbf{1}\left( \delta _{%
		\left[ i\right] }=1\right) }g_{i}\left( x\right) ^{B_{h\left[ i\right]
	}\left( x\right) \textbf{1}\left( \delta _{\left[ i\right] }=0\right) },\label{likelihood1}\tag{A7}
\end{align}
%\left( 1-\sum_{j=1}^{i-1}g_{j}\left(
%x\right) \right) ^{B_{h\left[ i\right] }\left( x\right) \bm{1}\left( \delta _{%
%		\left[ i\right] }=1\right) }g_{i}\left( x\right) ^{B_{h\left[ i\right]
%	}\left( x\right) \bm{1}\left( \delta _{\left[ i\right] }=0\right) }
where, for $i=1,\ldots,n$,  $q_{i}\left( x\right) =S_{0}(T_{(i)}^{-}| x)-S_{0}(T_{(i)}|x)$ are the
increments of $
S_{0}\left( t| x\right) $, and $g_{i}\left( x\right)
=G(T_{(i)}| x)-G(T_{(i)}^{-}| x)$ the increments of $G\left( t| x\right) $. Let  $P_{i}\left( x\right) =p\left( x\right) q_{i}\left( x\right) $ be the
increments of $S\left( t| x\right) $, then $\sum_{i=1}^{n}P_{i}\left(
x\right) =p\left( x\right) $. Maximizing (\ref{likelihood1}) is equivalent to maximizing the likelihood  
\begin{align}
L\left( X,T,\delta,\xi,\nu \right)  
=&\prod\limits_{i=1}^{n}P_{i}\left( x\right) ^{B_{h\left[ i\right]
	}\left( x\right) \textbf{1}\left( \delta _{\left[ i\right] }=1\right) }\left(
1-\sum_{j=1}^{i-1}P_{j}\left( x\right) \right) ^{B_{h\left[ i\right]
	}\left( x\right) \textbf{1}\left(\delta_{[i]}=0, \xi_{[i]}\nu _{[i]}=0\right)
}
\times \left( 1-\sum_{j=1}^{n}P_{j}\left( x\right) \right) ^{B_{h\left[ i%
		\right] }\left( x\right) \textbf{1}\left( \delta _{\left[ i\right] }=0,\xi_{[i]}\nu
	_{\lbrack i]}=1\right) }. \label{likelihood2}\tag{A8}
\end{align}
Further, consider the functions $\lambda _{i}\left( x\right) =P_{i}\left(
x\right) /\left\{ 1-\sum_{j=1}^{i-1}P_{j}\left( x\right) \right\} $
satisfying%
\begin{equation}
1-\sum_{j=1}^{k}P_{j}\left( x\right) =\prod_{j=1}^{k}\{1-\lambda _{j}\left(
x\right) \}.  \label{1-sumPj=prod(1-lambda_j)}\tag{A9}
\end{equation}%
Then, the increments $P_{i}\left( x\right) $ can be written in terms of $%
\lambda _{i}\left( x\right) $:%
\begin{equation}
P_{i}\left( x\right) =\lambda _{i}\left( x\right) \left\{
1-\sum_{j=1}^{i-1}P_{j}\left( x\right) \right\}=\lambda _{i}\left(
x\right) \prod_{j=1}^{i-1}\left\{1-\lambda _{j}\left( x\right) \right\}.
  \label{Pi=lambdai(1-sum)}\tag{A10}
\end{equation}
By substituting (\ref{1-sumPj=prod(1-lambda_j)}) and (\ref{Pi=lambdai(1-sum)}) in (\ref{likelihood2}), the likelihood (\ref{likelihood2}) is
\begin{align*}
L\left( X,T,\delta,\xi,\nu; p,S_{0}\right)  
=&\prod\limits_{i=1}^{n}\lambda _{i}\left( x\right) ^{B_{h\left[ i\right]
	}\left( x\right) \bm{1}\left( \delta _{\left[ i\right] }=1\right) }
\prod_{i=1}^{n}\left[ \prod_{j=1}^{i-1}\left\{1-\lambda _{j}\left( x\right)
\right\}\right] ^{B_{h\left[ i\right] }\left( x\right) \bm{1}\left( \delta _{\left[ i%
		\right] }=1\right) } \\
&\times \prod_{i=1}^{n}\left[ \prod_{j=1}^{i-1}\left\{1-\lambda _{j}\left( x\right)
\right\}\right] ^{B_{h\left[ i\right] }\left( x\right) \bm{1}\left( \delta _{%
		\left[ i\right] }=0,\xi _{\left[ i\right] }\nu _{\left[ i\right] }=0\right) }
\times \prod\limits_{i=1}^{n}\left[ \prod_{j=1}^{n}\left\{1-\lambda _{j}\left( x\right)
\right\}\right] ^{B_{h\left[ i\right] }\left( x\right) \bm{1}\left( \delta _{%
		\left[ i\right] }=0,\xi _{\lbrack i]}\nu _{\lbrack i]}=1\right) }.
\end{align*}
Taking into account that  $\prod_{i=1}^{n}\left[\prod_{j=1}^{i-1} a_j\right]^{b_i} = \prod_{i=1}^{n} a_i^{\sum_{j=i+1}^{n} b_j}$, where  $a_i$ and $b_i$, $i=1,\ldots,n$, are arbitrary sequences of nonnegative numbers, the likelihood becomes
 
\begin{align*}
L\left( X,T,\delta ,\xi,\nu; p,S_{0}\right)  
=&\prod_{i=1}^{n}\lambda _{i}\left( x\right) ^{B_{h\left[ i\right]
	}\left( x\right) \bm{1}\left( \delta _{\left[ i\right] }=1\right) }
\times\prod_{i=1}^{n}\left\{1-\lambda _{i}\left( x\right)
\right\}
^{\sum_{j=i+1}^{n}B_{h\left[ j\right] }\left( x\right) \bm{1}\left( \xi _{%
		\left[ j\right] }\nu _{\left[ j\right] }=0\right) +\sum_{j=1}^{n}B_{h\left[ j\right] }\left(
	x\right) \bm{1}\left(\delta_{[j]}=0,\xi _{\lbrack j]}\nu _{\lbrack
	j]}=1\right) 
}.
\end{align*}%
Maximizing the
likelihood $L\left( X,T,\delta,\xi,\nu; p,S_{0}\right) $ is equivalent to maximizing
the local log-likelihood: 
\begin{align*}
\Psi \{\lambda _{1}\left( x\right) ,\dots ,\lambda _{n}\left( x\right) \} 
=&\sum_{i=1}^{n}\left[ B_{h\left[ i\right] }\left( x\right) \bm{1}\left( \delta
_{\left[ i\right] }=1\right) \log \lambda _{i}\left( x\right) \right.  \\
&\left. +\left\{ \sum_{j=i+1}^{n}B_{h\left[ j\right] }\left( x\right)
\bm{1}\left( \xi _{\left[ j\right] }\nu _{\left[ j\right] }=0\right)
+\sum_{j=1}^{n}B_{h\left[ j\right] }\left(
x\right) \bm{1}\left( \delta_{[j]}=0,\xi _{\lbrack j]}\nu _{\lbrack
	j]}=1\right) \right\} \log \left( 1-\lambda _{i}
\right) \right] 
\end{align*}%
subject to $\prod_{i=1}^{n}\{1-\lambda _{i}\left( x\right)
\} =1-p\left( x\right) $. The maximizer $\lambda
_{i}\left( x\right) $ of the log-likelihood is 
\begin{align*}
\widehat{\lambda }_{i}\left( x\right) =&\frac{B_{h\left[ i\right] }\left( x\right) \bm{1}\left( \delta _{\left[ i%
		\right] }=1\right) }{\sum_{j=i}^{n}B_{h\left[ j\right] }\left(
	x\right) +\sum_{j=1}^{i-1}B_{h\left[ j\right] }\left( x\right)
	\bm{1}\left( \xi _{\lbrack j]}\nu _{\lbrack
		j]}=1\right) }.
\end{align*}%
In virtue of (\ref{Pi=lambdai(1-sum)}), the estimator $\widehat{S}_{h}^c(t| x)$  computed by forming the product of $\widehat{\lambda}_i$'s such that $T_{(i)} \leq t$ is the nonparametric maximum likelihood estimator of  $S(t| x)$. This completes the proof of Proposition \ref{prop:npmle}.

\noindent\textbf{Proof of Corollary \ref{coro:supFhat}}. The dominant part of $\widehat{\Lambda}_h^c(t| x)-\Lambda (t| x)$ in Theorem \ref{thm:lambda} verifies
	\begin{align}
 \sum_{i=1}^{n}B_{hi}(x)\zeta \left( T_{i},\delta _{i},\xi_{i},\nu _{i},t,x\right)  \notag 
=&\int_{0}^{t}\frac{d\widehat{H}_{h}^{1}\left( v\mid  x\right) }{J\left(
	v^-\mid  x\right) }-\int_{0}^{t}\frac{\widehat{J}_{h}\left( v^-\mid  x\right) }{%
	J^{2}\left( v^-\mid  x\right) }dH^{1}\left( v\mid  x\right)  \notag \\
	=&\int_{0}^{t}\frac{d\widehat{H}_{h}^{1}\left( v\mid  x\right)
	-dH^{1}(v\mid  x)}{J\left( v^-\mid  x\right) }-\int_{0}^{t}\frac{\widehat{J}%
	_{h}\left( v^-\mid  x\right) -J(v^-\mid  x)}{J^{2}\left( v^-\mid  x\right) }dH^{1}\left(
v\mid  x\right)  \notag \\
=&\left[ \frac{\widehat{H}_{h}^{1}\left( v\mid  x\right) -H^{1}\left( v\mid  x\right)
}{J\left( v^-\mid  x\right) }\right] _{0}^{t}+\int _{0}^{t}\frac{\widehat{H}%
	_{h}^{1}\left( v\mid  x\right) -H^{1}\left( v\mid  x\right) }{J^{2} \left( v^-\mid  x\right)}%
dJ(v\mid  x)  \notag \\
& -\int _{0}^{t}\frac{\widehat{J}_{h}\left( v^-\mid  x\right) -J(v^-\mid  x)}{%
	J^{2}\left( v^-\mid  x\right) }dH^{1}\left( v\mid  x\right)  \notag \\
\leq & \frac{1}{\theta }\sup_{a\leq t\leq b, x\in I}\mid \widehat{H}%
_{h}^{1}\left( t\mid  x\right) -H^{1}\left( t\mid  x\right) \mid +\frac{1}{%
	\theta }\sup_{a\leq t\leq b, x\in I}\mid \widehat{H}_{h}^{1}\left( t\mid  x\right)
-H^{1}\left( t\mid  x\right) \mid \notag \\
& -\frac{1}{\theta ^{2}}\sup_{a\leq t\leq b, x\in I}\mid \widehat{J}%
_{h}\left( t\mid  x\right) -J\left( t\mid  x\right) \mid .  \notag %\\
\end{align}
The last three terms in the inequality are bounded by applying Lemma 5 in \cite{iglesias1999strong}, which holds not only for conditional survival functions like $1-H(t| x)$, but also for conditional subdistribution functions  as $H^1(t| x)$ and $H^{11}(t| x)$ (see Remark 2 in \cite{iglesias1999strong} and the proof of Theorem 2.1 in \cite{dabrowska1989uniform}). As a consequence, the dominant term of $\widehat{\Lambda }_h^c(t| x) - \Lambda(t| x)$ is bounded by
	\begin{equation*}
	\sup_{a\leq t\leq b ,x\in I}\mid\sum\limits_{i=1}^{n}B_{hi}(x)\zeta \left( T_{i},\delta _{i},\xi _{i},\nu _{i},t,x\right) \mid =O\left\{ \left( nh\right)^{-1/2}\left(\log n\right)^{1/2}\right\}. 
	\end{equation*}
Using the results of Theorem \ref{thm:FhatFweigtht} it is straightforward to prove the second part of this corollary.

\noindent\textbf{Proof of Proposition \ref{prop:Bias_and_variance}}.	From Theorem \ref{thm:FhatFweigtht}, the bias of the nonparametric estimator $1-\widehat{F }^c_{h}\left( t| x\right) $ is asymptotically equal to the expected value 
	\begin{equation}
	\frac{(nh)^{-1}\left\{1-F\left( t\mid x\right)\right\}}{m\left(x\right)}\sum\limits_{i=1}^{n}K\left(\frac{x-X_{i}}{h}\right)\zeta\left(T_{i},\delta_{i},\xi_{i},\nu_{i},t,x\right)=I+II \label{eqn:I+II}\tag{A11}
	\end{equation}
	where
	\begin{align}
		I=&\frac{(nh)^{-1}\left\{1-F\left( t\mid x\right)\right\}}{m\left(x\right)}\Bigg[\sum\limits_{i=1}^{n}K\left(\frac{x-X_{i}}{h}\right) \zeta \left(T_{i},\delta_{i},\xi_{i},\nu_{i},t,x\right) -\text{E}\left\{ \sum_{i=1}^{n}K\left( \frac{x-X_{i}}{h}\right) \zeta \left( T_{i},\delta _{i},\xi _{i},\nu _{i},t,x\right) \right\}\Bigg], \label{eqn:I}\tag{A12}\\	%
		II=& \frac{(nh)^{-1}\{1-F\left( t\mid x\right)\}}{m\left(x\right)}\text{E}\left\{\sum\limits_{i=1}^{n}K\left( \frac{x-X_{i}}{h}\right) \zeta \left( T_{i},\delta _{i},\xi _{i},\nu _{i},t,x\right) \right\}.\label{eqn:II}\tag{A13}
		\end{align}	
	Since  $E(I)=0$, the asymptotic bias of the estimator $1-\widehat{F }_h^c\left(t\mid x\right)$ is $II$.
		Using Lemmas 1 and 2 in the Supplementary Material,  
	\begin{align*}
II	& =\frac{h^{2}\left\{1-F\left( t\mid x\right)\right\}(\Phi_c ^{\prime \prime }\left( x,t,x\right)m\left( x\right) +2\Phi_c ^{\prime }\left( x,t,x\right) m^{\prime }\left(x\right))d_{K} }{2m\left( x\right) }+O\left( h^{4}\right), 
	\end{align*}
	with $\Phi_c^\prime(y,t,x)$ and  $\Phi_c^{\prime\prime} (y,t,x)$  the first and  second derivatives of $\Phi_c \left( y,t,x\right)$ with respect to $y$. 
Recalling (\ref{eqn:I+II}),  the asymptotic
	variance of $1-\widehat{F}_{h}^c(t|x)$\ is  
	\begin{equation}
\var\left( I\right)=\frac{\left\{1-F\left( t\mid x\right)\right\} ^{2}}{m^{2}( x) } (V_1-V_2),  \label{eqn:(S)mVar(I)}\tag{A14}
\end{equation}%
where
\begin{align*}
V_1	=\frac{1}{nh^{2}}\text{E}\left\{ K^{2}\left( \frac{x-X}{h}\right) \zeta
^{2}\left( T,\delta,\xi,\nu,t,x\right) \right\}, \text{\ \ }
V_2	=\frac{1}{nh^{2}}\left[ \text{E}\left\{ K\left( \frac{x-X}{h}\right) \zeta
\left( T,\delta,\xi,\nu,t,x\right) \right\} \right]
^{2}.
\end{align*}%
From Lemmas 1 and 2 in the Supplementary Material,    $V_{2}$  reduces to
\begin{align}
V_{2}	&=\frac{1}{4}\frac{h^{2}}{n}d_{K}^{2}\left\{\frac{\Phi_c ^{\prime \prime }\left( x,t,x\right)m\left( x\right) +2\Phi_c ^{\prime }\left( x,t,x\right) m^{\prime }\left(x\right) }{m\left( x\right) }\right\}^{2}+O\left( \frac{h^{4}}{n}\right).  \label{V2}\tag{A15}
\end{align}
As for $V_{1}$, let us define $	\Phi _{1}^c\left( y,t,x\right) =E\left( \zeta ^{2}\left( T,\delta ,\xi ,	\nu ,t,x\right) | X=y\right).$ 	Then, after  a change of variable and  a Taylor's expansion (as in the proof of Lemma 1 in the Supplementary Material) we obtain
\begin{align}
V_{1} 	&=\frac{1}{nh}\Phi^c _{1}\left( x,t,x\right) m\left( x\right) c_{K}+\frac{1}{2
}\frac{h}{n}e_{K}\frac{d ^{2}}{d y^{2}}\left\{ \Phi^c_1\left( y,t,x\right)
m\left( y\right) \right\} |_{y=x}+O\left( n^{-1}h^{3}\right)\label{V1} \tag{A16}
\end{align}	
where $e_K=\int v^2 K^2(v)dv$. The proof concludes by substituting (\ref{V2}) and (\ref{V1})   into (\ref{eqn:(S)mVar(I)}).

%------------------------------------------------------

\noindent\textbf{Proof of Theorem \ref{thm:asymnormality}.}
From Theorem \ref{thm:FhatFweigtht}, we consider
\begin{equation*}
(nh)^{1/2}\left\{\widehat{F}_{h}^c \left( t\mid x\right) -F\left( t\mid x\right) \right\}=(nh)^{1/2}
 \left\{1- F\left(
t\mid x\right)\right\}\sum_{i=1}^{n}\widetilde{B}_{hi}\left( x\right)  \zeta \left( T_{i},\delta _{i},\xi _{i},\nu
_{i},t,x\right) +R_{n2}\left( t,x\right)
\end{equation*}%
\noindent with $\zeta(T, \delta, \xi, \nu, t,x)$ and $R_{n2}(t,x)$ given in (\ref{eqn:zeta}) and (\ref{eqn:Rn3}), respectively. 
The condition $ (\log n)^3 / nh \rightarrow 0$ implies that\\ $(nh)^{1/2}(\log n / nh)^{3/4} \rightarrow 0$, so the remainder term  $(nh)^{1/2}R_{n2}(t,x)$ is negligible. Consequently, the asymptotic distribution of $(nh)^{1/2}(\widehat{F}_{h}^c \left( t\mid x\right) -F\left( t\mid x\right) )$ is  that of
\begin{align}
(nh)^{1/2}\dfrac{1-F\left( t\mid x\right) }{m\left( x\right) }
\sum\limits_{i=1}^{n}\frac{1}{nh}K\left( \frac{x-X_{i}}{h}\right) \zeta
\left( T_{i},\delta _{i},\xi _{i},\nu _{i},t,x\right)  = (nh)^{1/2}(I + II),\label{eqn:I:II}\tag{A17}
\end{align}%
where $I$ and $II$ are given in (\ref{eqn:I}) and (\ref{eqn:II}). 
Under the assumption $nh^5\rightarrow 0$, we have $(nh)^{1/2}II=o(1)$. Therefore, the asymptotic distribution of (\ref{eqn:I:II}) is that of $(nh)^{1/2}I$. Let us define $(nh)^{1/2}I=\sum_{i=1}^{n}\eta_{i,h}(t,x)$,
where 
\begin{align*}
\eta_{i,h}(t,x) =&\dfrac{(nh)^{-1/2}\left\{1-F\left( t\mid x\right) \right\}}{m\left( x\right) }
\Bigg[ K\left( \frac{x-X_{i}}{h}\right) \zeta \left(
T_{i},\delta _{i},\xi _{i},\nu _{i},t,x\right)  -\text{E}\left\{ K\left( \frac{x-X_{i}}{h}\right) \zeta \left( T_{i},\delta
_{i},\xi _{i},\nu _{i},t,x\right) \right\} \Bigg],
\end{align*}%
 is a sequence of  $n$ independent random variables with mean  $0$. Note that
\begin{equation*}
\var(\eta_{i,h}(t,x)) = h\var(I) = \frac{1}{n}\frac{\left\{1-F(t \mid x)\right\}^2}{m(x)} \Phi_1^c(x,t,x)c_K + O\left(\frac{h^2}{n}\right)  = \frac{1}{n}s^2_c(t,x) + O\left(\frac{h^2}{n}\right) 
\end{equation*}
with $\var(I)$ in (\ref{eqn:(S)mVar(I)}) and $s_c^2(t,x)$ in (\ref{def:s_2}). Since $\var(\eta_{i,h}(t,x))<\infty$ for $i=1,\ldots,n$ and  $\var(\eta_{h}(t,x))=\sum_{i=1}^{n}\var(\eta_{i,h}(t,x))$ is positive, then we can apply  Lindeberg's theorem  \citep{Billingsley1968} to obtain
 %[Theorem 7.2 in ][p. 42]
\begin{equation*}
\frac{\sum_{i=i}^{n}\eta _{i,h}\left( t,x\right) }{s_{c}^{2}(t,x)}\rightarrow N\left( 0,1\right) \text{\ \ in distribution}.
\end{equation*}
 Therefore, $(nh)^{1/2}\left\{ \widehat{F}_h^c(t| x)-F(t| x)\right\} \rightarrow
 N(0,s_{c}^{2}(t,x))$ in distribution.   	This proves (i). 
 In parallel to the  proof (i) we can prove (ii) as follows, note that		
 if $nh^5=C^5$ then the bias term is $(nh)^{1/2}II = (nh)^{1/2}\{h^2B_c(t,x) + O(h^4)\} = (nh^5)^{1/2}B_c(t,x) + O\{(nh^9)^{1/2}\} $ with $B_c(t,x)$ in (\ref{def:Bc}). 
Thus,  $(nh)^{1/2}\left\{ \widehat{F}_h^c(t| x)-F(t| x)\right\} \rightarrow
 N( C^{5/2}B_c(t,x),s_{c}^{2}(t,x))$ in distribution. This completes the proof.

\bibliography{wileyNJD-APA}%

\end{document}